%% file: Model2_v2.tex
\documentclass[%
superscriptaddress,
nofootinbib,
 amsmath,amssymb,
 aps,
pra,
floatfix,
]{revtex4-2}
\usepackage{array,multirow,graphicx,makecell}
\usepackage[font=small,labelfont=bf,justification=justified]{caption}
\usepackage{subcaption}
\usepackage{ftnxtra}
\usepackage{booktabs}
\usepackage{dcolumn}
\usepackage{hhline}
\usepackage{bm}
\usepackage[colorlinks,citecolor=blue,urlcolor=blue,linkcolor=blue]{hyperref}
\usepackage{color}
\usepackage{orcidlink}  
\usepackage{graphicx}
\usepackage{physics}
\usepackage{tabularray}
\usepackage{threeparttable}
\usepackage{comment}
\usepackage{float}
\usepackage{bm}
\usepackage{natbib}

\renewcommand{\arraystretch}{1.5}

\graphicspath{{./}{Fig/}}

\usepackage{xcolor}

\begin{document}


\title{Testing Extended Exponential Infrared Teleparallel Gravity with CMB, DESI, and PantheonPlus}

\author{Mahmoud Hashim\,\orcidlink{0000-0003-0691-9264}}\email{mahmoud.hashim@bue.edu.eg}
\affiliation{Centre for Theoretical Physics, The British University in Egypt, P.O. Box 43, El Sherouk City, Cairo 11837, Egypt}
\author{Jackson Levi Said\,\orcidlink{0000-0002-7835-4365}}\email{jsaid01@um.edu.mt}
\affiliation{Institute of Space Sciences and Astronomy, University of Malta, Malta, MSD 2080 Department of Physics, University of Malta, Malta}
\author{Waleed El Hanafy\,\orcidlink{0000-0002-0097-6412}}\email{waleed.elhanafy@bue.edu.eg}
\affiliation{Centre for Theoretical Physics, The British University in Egypt, P.O. Box 43, El Sherouk City, Cairo 11837, Egypt}
\author{Eleonora Di Valentino\,\orcidlink{0000-0001-8408-6961}}\email{e.divalentino@sheffield.ac.uk}
\affiliation{School of Mathematical and Physical Sciences, University of Sheffield, Hounsfield Road, Sheffield S3 7RH, United Kingdom}


\begin{abstract}
We place cosmological constraints on an extended exponential infrared (EExpIR) $f(T)$ teleparallel gravity model using current observational datasets. The extension is characterized by an additional parameter, $b$, which governs the late-time cosmological dynamics. Owing to the existence of two solution branches, phase-space analysis is employed to select the cosmological branch and to establish an appropriate prior range for the new parameter. In this framework, the torsion sector acts as an effective dark-energy component, exhibiting a smooth evolution from a quintessence-like phase to a phantom regime, while recovering the standard cosmological behavior at early times. To constrain the model parameters, we combine early-Universe observations from Planck, ACT, and SPT measurements of the CMB with late-time probes from DESI BAO and PantheonPlus Type Ia supernovae. Using the full dataset, including the SH0ES calibration, we obtain $b=0.573^{+0.016}_{-0.060}$ and $H_0=70.67^{+0.31}_{-0.35}~{\rm km\,s^{-1}\,Mpc^{-1}}$ at 68\% CL. Model comparison based on Bayesian evidence reveals that the EExpIR $f(T)$ and the $\Lambda$CDM models are statistically indistinguishable when only CMB data are considered, whereas the inclusion of the full cosmological dataset with SH0ES leads to a weak preference in favor of $\Lambda$CDM.
\end{abstract}

\maketitle


\section{Introduction}

The discovery that the Universe is undergoing an accelerated expansion
constitutes one of the most profound challenges in modern
cosmology~\cite{Riess:1998cb,Perlmutter:1998np}.
Within the standard cosmological model, $\Lambda$CDM, evidence from
Type Ia supernovae, cosmic microwave background (CMB) measurements,
and large-scale structure observations indicates that approximately
70\% of the present cosmic energy budget is associated with a component
possessing negative pressure, identified with a cosmological constant
$\Lambda$~\cite{Planck:2018vyg}.
Despite its remarkable observational success, the cosmological constant
interpretation is affected by well-known theoretical difficulties,
including the fine-tuning and coincidence
problems~\cite{Weinberg:1988cp,Zlatev:1998tr}.
These challenges have motivated extensive efforts to explore alternative
explanations for cosmic acceleration.

One possibility is that the late-time acceleration arises from a
dynamical dark-energy sector~\cite{Copeland:2006wr}.
Another appealing avenue is to modify the gravitational interaction
itself on cosmological scales~\cite{Clifton:2011jh}.
In modified gravity theories, the observed acceleration can be interpreted
as a geometric effect rather than the result of an additional cosmic fluid.
Among the various modified gravity frameworks, teleparallel gravity has
attracted considerable attention in recent
years~\cite{Bahamonde:2021gfp}.
Unlike General Relativity (GR), where gravitation is described by spacetime
curvature through the Levi-Civita connection, teleparallel gravity is
formulated in terms of spacetime torsion using a curvature-free, metric-compatible connection.
The teleparallel equivalent of General Relativity (TEGR) yields the same
classical field equations as GR, while providing a distinct geometrical
description of gravity~\cite{Bahamonde:2021gfp}.

A natural extension of TEGR is obtained by replacing the torsion scalar
$T$ in the gravitational action with an arbitrary function $f(T)$.
The resulting $f(T)$ theories have been extensively investigated as
gravitational alternatives to dark
energy~\cite{Bengochea:2008gz,Linder:2010py}.
An attractive property of these models is that their field equations
remain second order, which can simplify cosmological
analyses~\cite{Linder:2010py}.
Depending on the particular functional form of $f(T)$, the effective
torsion sector can mimic quintessence, phantom dark energy, or even a
transition across the phantom divide without introducing a separate
dark-energy scalar field~\cite{Bahamonde:2021gfp}.
Among the numerous forms proposed in the literature, exponential
infrared modifications are particularly interesting candidates.
For positive $b$ in the model considered below, these modifications
become negligible relative to the TEGR term at high redshift, while
becoming significant at late times.
Consequently, they can produce the present accelerated expansion through
purely gravitational effects while recovering the standard early-time
background evolution asymptotically.

In this work we focus on the Extended Exponential Infrared (EExpIR)
model of $f(T)$ gravity, defined by
\[
f(T)=T\,e^{\beta\left(T_0/T\right)^b},
\]
where $T_0=6H_0^2$ is the present value of the torsion scalar in a
spatially flat cosmological background, and $b$ and $\beta$ are
dimensionless parameters.
The present-day Friedmann equation fixes $\beta$ in terms of $b$ and
the present matter and radiation density parameters, once a solution
branch is selected~\cite{Santos:2022atq}.
The model therefore introduces one additional free cosmological
parameter relative to $\Lambda$CDM.
It does not reproduce $\Lambda$CDM exactly for any particular value
of $b$, although it reduces to TEGR, and hence GR without a cosmological
constant, when $\beta=0$.
On the positive-$\beta$ branch considered here, the effective dark-energy
sector can exhibit qualitatively different behaviors:
(i) phantom-like evolution for $b\geq1$,
(ii) a quintessence-to-phantom transition for $0.5<b<1$, and
(iii) a transition from negative to positive effective dark-energy
density for $0<b<0.5$~\cite{Santos:2022atq}.
Previous analyses combining background and structure-growth observations
obtained central values spanning approximately $b\simeq0.67$--$0.87$,
depending on the data combination, suggesting that the model can realize
a dynamical effective dark-energy
evolution~\cite{Santos:2022atq}.

An additional feature of the EExpIR model is the existence of multiple
solution branches associated with the Lambert-$W$ representation of the
present-day Friedmann constraint.
This structure allows for distinct cosmological evolutions, making it
important to determine which branches admit a physically acceptable
background history.
For this reason, we first perform a detailed phase-space analysis before
confronting the model with cosmological data.

The increasing precision of observational cosmology now permits stringent
tests of modified gravity scenarios.
In particular, measurements of the CMB anisotropies constrain both the
background expansion and the evolution of linear
perturbations~\cite{Planck:2018vyg}, while baryon acoustic oscillation
(BAO) measurements and the Pantheon+ Type Ia supernova compilation
provide powerful probes of the late-time
Universe~\cite{Scolnic:2021amr}.
Combining these datasets enables a comprehensive assessment of the
observational viability of alternative cosmological models.

The primary goal of the present work is to perform a comprehensive
observational analysis of the EExpIR model at both the background and
linear perturbation levels.
We first identify cosmological branches admitting viable background
evolution through dynamical systems methods and then constrain the model
using current cosmological observations, including CMB, DESI BAO, and
PantheonPlus data.
We further investigate the implications of the model for cosmological
parameter estimation, study the evolution of the effective dark-energy
sector, and assess its statistical performance relative to $\Lambda$CDM
through Bayesian model comparison.

This work is carried out within the Cosmology Compilation Group
initiative as part of the CosmoVerse
project~\cite{CosmoVerseNetwork:2025alb}.
The paper is organized as follows.
In Sec.~\ref{Sec:fT_gravity}, we review teleparallel geometry and the
$f(T)$ field equations.
In Sec.~\ref{Sec:fT_cosmology}, we present the modified Friedmann
equations and the linear perturbation equations.
In Sec.~\ref{Sec:expfT_family}, we introduce a family of exponential
$f(T)$ theories, discuss the different branches of the EExpIR solution,
and examine their phase portraits for different parameter values.
In Sec.~\ref{Sec:data}, we describe the observational datasets and
statistical methodology employed in the parameter estimation analysis.
In Sec.~\ref{Sec:results}, we present the resulting cosmological
constraints and model-selection results.
Finally, Sec.~\ref{Sec:summary} summarizes our main findings and
conclusions.

\section{Teleparallel Geometrical and \texorpdfstring{$f(T)$}{} Gravity}\label{Sec:fT_gravity}

This section provides a concise overview of teleparallel geometry, highlighting its fundamental geometrical structures and key relations. Subsequently, we present the modified gravitational field equations arising within the framework of $f(T)$ gravity.

\subsection{The Geometric Description}

We begin with a brief introduction to teleparallel geometry. A teleparallel spacetime can be described by the pair ($M$, $e_a$) in the Weitzenb\"ock gauge, where $M$ is an $n$-dimensional smooth manifold and $e_a\, (a=1,\cdots,n)$ are $n$ linearly independent vector (vielbein) fields defined on it. Denoting the components of these vector fields with respect to a coordinate basis $\partial_\mu$ by $e_a{^\mu}\, (\mu=1,\cdots,n)$, one can introduce the dual set of independent one-forms $e^a=e^a{_\mu}dx^\mu$, satisfying\footnote{Both the Greek (coordinate) and the Latin (Lorentz) indices are subject to the Einstein summation convention.}
\begin{equation}\label{eq:vielbein}
    e_a{^\mu}e^a{_\nu}=\delta^\mu_\nu, \qquad e_a{^\mu}e^b{_\mu}=\delta^b_a.
\end{equation}
where $\delta$ denotes the Kronecker delta. In a general Lorentz gauge, the teleparallel connection is given by~\cite{Krssak:2015oua}
\begin{equation}\label{eq:W_connction}
\Gamma{^\alpha}{_{\mu \nu}}=e_a{^\alpha} \left( \partial_\mu e^a{_\nu}+\omega{^a}{_{b\mu}} e{^b}{_\nu} \right),
\end{equation}
where $\omega{^a}{_{b\mu}}$ denotes the flat inertial spin connection, with respect to which the vielbein fields satisfy the following postulate
\begin{equation}\label{eq:W_Cov_derv}
    \partial_\mu e^a{_\nu}+\omega{^a}{_{b\mu}} e{^b}{_\nu}-\Gamma{^\alpha}_{\mu \nu} e^a{_\alpha}=0.
\end{equation}
For a given vielbein and spin connection, this postulate uniquely determines the affine connection $\Gamma$.
Here, $\nabla$ denotes the covariant derivative associated with the
teleparallel connection, $\nabla{_\mu} V{^\alpha}=\partial{_\mu} V{^\alpha} + \Gamma{^\alpha}{_{\mu\nu}}V{^\nu}$.

The commutator of covariant derivatives acting on an arbitrary vector field $V_a$ can be expressed as
\begin{equation}\label{eq:comm_rel}
\nabla_{\nu}\nabla_{\mu}V_{a}{^{\alpha}} - \nabla_{\mu}\nabla_{\nu}V_{a}{^{\alpha}} = -R^{\alpha}{_{\epsilon\mu\nu}}
V_{a}{^{\epsilon}} + T^{\epsilon}{_{\mu\nu}} \nabla_{\epsilon} V_{a}{^{\alpha}}.
\end{equation}
The curvature tensor associated with the teleparallel connection vanishes,
\[
R^{a}{_{b\mu\nu}}= \partial{_\mu}\omega{^a}{_{b\nu}}-\partial{_\nu}\omega{^a}{_{b\mu}}+\omega{^a}{_{c\mu}}\omega{^c}{_{b\nu}}-\omega{^a}{_{c\nu}}\omega{^c}{_{b\mu}}=0,
\]
whereas the torsion tensor is given by
\begin{equation}\label{eq:Torsion}
T{^\alpha}_{\mu \nu}=\Gamma{^\alpha}_{\mu \nu}-\Gamma{^\alpha}_{\nu \mu}=e{_a}{^\alpha} \left(\partial_\mu e{^a}{_\nu}-\partial_\nu e{^a}{_\mu}+\omega{^a}{_{b \mu}} e{^b}{_\nu}-\omega{^a}{_{b \nu}} e{^b}{_\mu}\right),
\end{equation}
which is antisymmetric in its last two indices by construction. The contortion tensor is defined as
\begin{equation}\label{eq:Contortion}
K{^\alpha}_{\mu \nu}=\frac{1}{2}\left(T{_\nu}{^\alpha}{_\mu}+T{_\mu}{^\alpha}{_\nu}-T{^\alpha}{_{\mu \nu}}\right),
\end{equation}
and is antisymmetric in the first two indices when all indices are lowered. Using the contortion tensor, one introduces the superpotential tensor
\begin{equation}\label{eq:SuperPot}
S^{\alpha\mu\nu}=\frac{1}{2}\left(K^{\mu \nu \alpha}+g^{\alpha \mu}T{^{\sigma \nu}{_\sigma}}-g^{\alpha \nu}T{^{\sigma \mu}{_\sigma}}\right),
\end{equation}
which is antisymmetric in the last pair of indices. Within teleparallel geometry, a fundamental invariant known as the torsion scalar is then defined through
\begin{eqnarray}\label{eq:T_scalar}
T&=&S{_\alpha}{^{\mu \nu}}T{^\alpha}{_{\mu \nu}},\nonumber\\
 &=&\frac{1}{4}T^{\alpha \mu \nu}T_{\alpha \mu \nu}+\frac{1}{2}T^{\alpha \mu \nu}T_{\nu \mu \alpha}-T{_\mu} T{^\mu},
\end{eqnarray}
where $T_\mu=T{^\alpha}{_{\mu\alpha}}=-T{^\alpha}{_{\alpha\mu}}$ denotes the torsion vector.

A pseudo-Riemannian metric on $M$ can also be constructed from the tetrad fields according to
\begin{equation}\label{eq:metric}
g_{\mu \nu} := \eta_{ab}e^{a}{_{\mu}}e^{b}{_{\nu}},
\end{equation}
whose inverse takes the form
\begin{equation}\label{eq:inverse}
g^{\mu \nu} = \eta^{ab}e_{a}{^{\mu}}e_{b}{^{\nu}},
\end{equation}
with $\eta_{ab}$ representing the Minkowski metric in the tangent space with signature $(-,+,+,+)$. The corresponding Levi-Civita connection is given by
\begin{equation}\label{eq:Christoffel}
\mathring{\Gamma}{^{\alpha}}{_{\mu\nu}}:= \frac{1}{2} g^{\alpha \sigma}\left(\partial_{\nu}g_{\mu \sigma}+\partial_{\mu}g_{\nu \sigma}-\partial_{\sigma}g_{\mu \nu}\right).
\end{equation}

It is to be noted that the spin connection $\omega{^a}{_{b\mu}}$ is purely inertial, so in an appropriate frame it can be set to zero locally, defining the Weitzenb\"{o}ck gauge~\cite{Bahamonde:2021gfp}. Both the Levi-Civita and Weitzenb\"{o}ck connections satisfy the metricity condition~\cite{Weitzenbock:1923efa,Maluf:2013gaa},
\begin{equation}\label{eq:metricity}
    \mathring{\nabla}_{\alpha}g_{\mu \nu}=0={\nabla}_{\alpha}g_{\mu \nu},
\end{equation}
where $\mathring{\nabla}$ denotes the covariant derivative associated with the Levi-Civita connection. Furthermore, one finds that the Ricci scalar constructed from the Levi-Civita connection, $\mathring{R}$, differs from the negative of the torsion scalar only by a total divergence term,
\begin{equation}\label{eq:Ricci_sc}
    \mathring{R}=-T-2\mathring{\nabla}{_\mu} T{^\mu}.
\end{equation}
As a consequence, $-T$ may replace the Ricci scalar in the Einstein-Hilbert action without altering the resulting field equations. This leads to the Teleparallel Equivalent of General Relativity (TEGR), which is dynamically equivalent to GR. In the following we restrict our treatment to the pure tetrad formalism in 4-dimensional spacetime, using an appropriate tetrad compatible with a vanishing spin connection. For more details about teleparallel gravity and its applications, see the review~\cite{Bahamonde:2021gfp}.

\subsection{The \texorpdfstring{$f(T)$}{} Field Equations}

Among the most widely studied extensions of TEGR is $f(T)$ gravity, in which the torsion scalar appearing in the action is generalized to an arbitrary function $f(T)$. The corresponding action is
\begin{equation}\label{eq:action}
    \mathcal{S}=-\frac{1}{2\kappa^2}\int d^4 x\, e f(T)+\int d^4 x\,\mathcal{L}_m.
\end{equation}

Here, $\kappa$ is related to Newton's gravitational constant through $\kappa^2=8\pi G_N/c^4$, $\mathcal{L}_m$ denotes the matter Lagrangian density, and $e$ is the determinant of the tetrad field $e^{a}_{\mu}$, where $f(T)=T$ gives the TEGR limit, which is dynamically equivalent to GR. It is straightforward to verify that $e=\sqrt{-g}$, where $g$ is the determinant of the metric tensor.

Varying the action \eqref{eq:action} with respect to the tetrad field yields the field equations
\begin{equation}\label{eq:field_eqn}
    f_T\mathfrak{G}_{\mu \nu}-\frac{1}{2} g_{\mu \nu}\left(T\, f_T - f\right)+2 S_{\mu \nu}{^\sigma}\, \partial_\sigma T f_{TT}=\kappa^2 \mathcal{T}_{\mu \nu},
\end{equation}
where $\mathfrak{G}_{\mu \nu}$ is the Einstein tensor, $\mathcal{T}_{\mu \nu}$ denotes the matter energy-momentum tensor, and $f_T\equiv \frac{df(T)}{dT}$ and $f_{TT}\equiv \frac{d^2f(T)}{dT^2}$. It should be emphasized that, although TEGR preserves local Lorentz invariance at the level of the field equations, this property is generally lost in the pure vielbein formalism of $f(T)$ gravity~\cite{Li:2010cg}.

For cosmological applications, it is often convenient to rewrite the field equations in the form
\begin{equation}\label{eq:EoM}
    \mathfrak{G}_{\mu \nu}=\kappa^2\left(\mathcal{T}_{\mu\nu}+\mathfrak{T}_{\mu\nu}\right),
\end{equation}
where $\mathcal{T}_{\mu\nu}$ represents the matter stress-energy tensor, while $\mathfrak{T}_{\mu\nu}$ corresponds to an effective torsional stress-energy tensor generated by the $f(T)$ corrections. In this representation, the torsional sector can effectively reproduce the role usually attributed to dark energy. The explicit expression for $\mathfrak{T}_{\mu\nu}$ is
\begin{equation}\label{eq:torsion_Tmn}
    \mathfrak{T}_{\mu\nu}=\frac{1}{\kappa^2}\left[\frac{1}{2}g_{\mu\nu}(T f_T-f)-2S{_{\mu\nu}}{^\sigma} \partial_\sigma T f_{TT}+(1-f_T)\mathfrak{G}_{\mu\nu}\right].
\end{equation}

Since this effective energy-momentum tensor originates entirely from geometric quantities, the resulting cosmic acceleration can be interpreted as a consequence of spacetime geometry itself, thereby avoiding the introduction of a fundamental dark energy fluid.

\section{The \texorpdfstring{$f(T)$}{} Cosmology}\label{Sec:fT_cosmology}

A comprehensive assessment of modified gravity (MG) theories requires examining both the background evolution of the Universe and the dynamics of cosmological perturbations. At the background level, the expansion history is characterized by the Hubble parameter $H(z)$, which is constrained by geometrical probes such as Type Ia supernovae (SNIa) and BAO. Consequently, the background evolution determines the effective dark energy behavior and the overall expansion dynamics. Nevertheless, distinct gravitational theories may reproduce nearly indistinguishable expansion histories while predicting different behaviors for the formation and evolution of cosmic structures.

This degeneracy can be reduced by considering linear perturbations. The evolution of matter density contrasts and gravitational potentials directly affects a wide range of observables, including galaxy clustering, weak gravitational lensing, redshift-space distortions, and CMB anisotropies. In general, modifications to gravity alter the growth of matter perturbations, modify the effective gravitational coupling, and may introduce a difference between the metric potentials, commonly referred to as the gravitational slip. Such effects can lead to observable departures from the predictions of GR.

Therefore, a simultaneous investigation of the background expansion and linear perturbation evolution provides a robust and complementary approach for testing cosmological models. This framework is particularly useful for distinguishing geometric extensions of gravity, such as $f(T)$ theories, from the standard $\Lambda$CDM cosmology. In the following subsections, we summarize the main equations governing the background and linear perturbation evolution in $f(T)$ cosmology.

\subsection{The \texorpdfstring{$f(T)$}{} Background Evolution}

To study the cosmological background dynamics, we assume a homogeneous and isotropic Universe described by the spatially flat Friedmann-Lema\^{\i}tre-Robertson-Walker (FLRW) line element
\begin{equation}\label{eq:FRW}
    dS^2 = - dt^2 + a^2(t) \delta_{ij} dx^i dx^j,
\end{equation}
where $a(t)$ denotes the cosmic scale factor. Throughout this work, natural units with $c=1$ are employed. The metric \eqref{eq:FRW} is generated by the diagonal tetrad
\begin{equation}\label{eq:FRW_tetrad}
    e{_\mu}{^a} = \mathrm{diag}(1, a(t), a(t), a(t)),
\end{equation}
through the relation \eqref{eq:metric}. We note that the above setting is consistent with an identically vanishing spin connection $\omega{^a}{_{b\mu}}\equiv 0$. Substituting this tetrad into the torsional quantities of teleparallel gravity leads to the simple expression
\begin{equation}\label{eq:FRW_Tsc}
    T=6 H^2,
\end{equation}
where $H=\dot{a}/a$ is the Hubble parameter.

Assuming the matter content of the Universe can be modeled as a perfect fluid, the corresponding energy-momentum tensor takes the form
\begin{equation}\label{eq:matter_Tmn}
    \mathcal{T}_{\mu\nu}=\rho u_\mu u_\nu + p (u_\mu u_\nu + g_{\mu \nu}),
\end{equation}
with $\rho$ and $p$ representing the energy density and pressure of the fluid, respectively, while $u^{\mu}=dx^{\mu}/\sqrt{-ds^2}$ is its normalized four-velocity. Inserting the FLRW tetrad into the field equations \eqref{eq:field_eqn} yields the modified Friedmann equations
\begin{eqnarray}
  3H^2 &=& \kappa^2 (\rho +\rho_T), \label{eq:fT_FR1}\\
  -2\dot{H}-3H^2 &=& \kappa^2 (p+p_T),\label{eq:fT_FR2}
\end{eqnarray}
where the effective torsional energy density and pressure are given by
\begin{eqnarray}
       \rho_T &=& \frac{1}{2\kappa^2}\left(T+f-2Tf_T\right),\label{eq:T_dens} \\
       p_T &=& -\frac{1}{2\kappa^2}\left(T+f-2Tf_T\right) -\frac{2}{\kappa^2}\dot{H}\left(1-f_T-2T f_{TT}\right).\label{eq:T_press}
\end{eqnarray}

Considering a matter sector obeying the linear barotropic equation of state $p(\rho)=w\rho$, with constant $w$, the cosmological dynamics can be reduced to the single evolution equation~\cite{Awad:2017yod}
\begin{equation}\label{eq:Hdot}
    \dot{H}=(1+w)\frac{f-2T f_T}{4(f_T+2T f_{TT})}.
\end{equation}

A particularly important feature follows from the relation \eqref{eq:FRW_Tsc}, which allows the right-hand side of Eq.~\eqref{eq:Hdot} to be expressed solely in terms of $H$. As a result, the cosmological evolution is described by a one-dimensional autonomous dynamical system, providing a useful framework for phase-space investigations.

In analogy with dark energy phenomenology, one may define an effective equation-of-state parameter associated with the torsional sector as
\begin{equation}\label{eq:T_EoS}
    w_T=\frac{p_T}{\rho_T}=-1-(1+w)\frac{(f-2T f_T)(1-f_T-2T f_{TT})}{(f_T+2T f_{TT})(T+f-2T f_T)}.
\end{equation}
This quantity characterizes the effective dark energy behavior induced by the modified gravitational sector and enables a direct comparison with conventional dark energy models.

For observational applications, including separately conserved pressureless matter and radiation, it is convenient to rewrite the modified Friedmann equation in terms of the redshift $z$. Introducing the normalized Hubble parameter $E(z)=H(z)/H_0$, one obtains
\begin{equation}\label{eq:E_FR}
    E^2=\Omega_\text{m}(1+z)^3+\Omega_\text{r}(1+z)^4+\Omega_\text{T} y(T),
\end{equation}
where $\Omega_\text{m}$, $\Omega_\text{r}$, and $\Omega_\text{T}$ denote the present-day density parameters of matter, radiation, and the torsional component, respectively. Assuming spatial flatness, these parameters satisfy $\Omega_\text{T}=1-\Omega_\text{m}-\Omega_\text{r}$. The function $y(T)$ represents the distortion function that encodes deviations from the standard cosmological evolution and is defined as
\begin{equation}\label{eq:distortion}
    y(T)=\frac{\kappa^2}{3\Omega_\text{T}H_0^2} \rho_T=\frac{1}{6\Omega_\text{T}H_0^2}(T+f-2Tf_T).
\end{equation}

The distortion function provides a convenient measure of the impact of modified teleparallel gravity on the cosmic expansion history. In particular, the choice $f(T)=T$ leads to $\rho_T=0$, recovering GR and the TEGR limit. Similarly, the linear form $f(T)=T+2\Lambda$ yields $y(T)=1$, reproducing the background evolution of the standard $\Lambda$CDM model. More generally, nonlinear forms of $f(T)$ can generate an effective time-dependent dark energy behavior without introducing an explicit dark energy fluid. Instead, accelerated expansion can emerge from the geometrical modifications encoded in the torsional sector. Within this framework, the usual matter components, including baryons, radiation, and cold dark matter, retain their standard conservation equations, while the role of dark energy is effectively replaced by the $f(T)$ gravitational contribution. We refer to such models as $f$CDM cosmologies.

\subsection{The \texorpdfstring{$f(T)$}{} Linear Perturbation Evolution}

While the background evolution determines the overall expansion history of the Universe, the formation and evolution of cosmological structures are governed by perturbations around this homogeneous background. The study of linear perturbations is therefore essential for connecting $f(T)$ gravity with observations such as galaxy clustering, weak lensing, redshift-space distortions, and CMB anisotropies.

For the perturbative analysis, it is convenient to introduce the conformal time $\tau$ through the relation $d\tau=dt/a(t)$. In terms of conformal time, the FLRW metric \eqref{eq:FRW} becomes
\begin{equation}
dS^2=a^2(\tau)\left[-d\tau^2+\delta_{ij}dx^i dx^j\right],
\end{equation}
and can be generated by the background vierbein
${e_{\mu}}^{a}=a(\tau) \, \text{diag}\left(1,1,1,1\right)$, see~\cite{Golovnev:2018wbh}. At the linear level, scalar perturbations are introduced through the Newtonian gauge line element
\begin{equation}
dS^2=a^2(\tau)\left[-(1+2\psi)d\tau^2+(1-2\phi)\delta_{ij}dx^i dx^j\right],
\end{equation}
where the scalar potentials $\psi$ and $\phi$ represent fluctuations of the temporal and spatial components of the metric, respectively. These potentials play a central role in determining the motion of matter and light in the perturbed Universe.

In teleparallel gravity the tetrad, rather than the metric, constitutes the fundamental gravitational variable. Consequently, linear perturbations contain additional components associated with the tetrad degrees of freedom. The most general scalar and pseudoscalar perturbations compatible with the metric perturbations above can be expressed as~\cite{Golovnev:2018wbh}
\begin{equation}
    e^0_0  =  a(\tau)\left(1+\psi\right), \,
    e^0_i  =  a(\tau) \partial_i \zeta,\,
    e^a_0  =  a(\tau) \partial_a \zeta,\,\text{and }
    e^a_j  =  a(\tau) \left((1-\phi)\delta^a_j+\epsilon_{ajk}\partial_k s\right),
\end{equation}
where $a,i,j,k=1,2,3$ in this equation, $\zeta$ is associated with scalar Lorentz boosts, and $s$ is the pseudoscalar associated with spatial rotations. Vector perturbations arise from divergence-free vector fields and will not be considered in the present work. Since the Newtonian gauge is already imposed, the scalar perturbation $\zeta$ appears symmetrically in both mixed components of the tetrad.

A distinctive feature of teleparallel theories is the existence of tetrad components that are not directly encoded in the metric. Besides the ten metric components familiar from GR, the tetrad contains six additional variables related to local Lorentz transformations. These extra components split into boost and rotational sectors. Nevertheless, the theory preserves diffeomorphism invariance, allowing coordinate gauge freedom to be fixed independently of these Lorentz components~\cite{Golovnev:2018wbh}.

At linear order around the FLRW background considered here, the rotational pseudoscalar mode $s$ decouples from the cosmological field equations. As a result, the scalar sector entering these equations consists of the two Newtonian potentials together with a single additional variable, $\zeta$, which is constrained by the antisymmetric part of the gravitational field equations.

Vector perturbations typically decay with the cosmic expansion in the absence of sources and make only a negligible contribution to the large-scale cosmological observables considered here. Tensor perturbations are gauge invariant at linear order. Therefore, in what follows we focus on the scalar perturbation sector, which carries most of the relevant information regarding departures from GR.

The perturbed field equations naturally separate into antisymmetric and symmetric sectors.

\subsubsection{Antisymmetric part}

The antisymmetric equations encode the constraint associated with the extra tetrad degree of freedom.

\begin{itemize}

\item The spatial components: Those are automatically satisfied, leaving $s$ unconstrained.

\item The mixed components:

\begin{equation}\label{asymm}
k^2\zeta=3\left(\phi^{\prime}+{\mathcal H}\psi-\frac{{\mathcal H}^{\prime}-{\mathcal H}^2}{\mathcal H}\phi\right),
\end{equation}

where ${\mathcal H}=aH$ denotes the conformal Hubble parameter, a prime indicates differentiation with respect to conformal time, and $k$ is the comoving wavenumber. This relation determines the extra scalar mode $\zeta$ in terms of the metric perturbations and background expansion. Consequently, the evolution of $\zeta$ is not independent but is sourced by the scalar gravitational potentials themselves.

\end{itemize}

\subsubsection{Symmetric part}

The symmetric sector governs the evolution of density perturbations and gravitational potentials. Its structure closely resembles the perturbation equations of standard cosmology, although important modifications arise through the effective gravitational coupling and the additional scalar variable.

\begin{itemize}

\item The temporal component\footnote{In what follows, departures from the standard cosmological scenario are quantified through the parameters $Q=\frac{1}{f_T}$ and $\Xi=12(\mathcal{H}'-\mathcal{H}^2)Qf_{TT}$, with the GR limit recovered for $Q\rightarrow 1$ and $\Xi\rightarrow 0$.}:

\begin{equation}
\label{eq:pert1}
k^2\phi+3{\mathcal H}(\phi^{\prime}+{\mathcal H}\psi)+\frac{3\mathcal{H}^2 \Xi}{a^2}\phi=-4\pi Q G_N a^2\delta\rho,
\end{equation}

where $\delta\rho$ denotes the matter density perturbation. The variable $\zeta$ has been eliminated using Eq.~\eqref{asymm}. Compared with GR, the modification is encoded through the functions $Q$ and $\Xi$, which alter the response of the metric potential to matter inhomogeneities.

\item The mixed components:

\begin{equation}
\label{smix}
\phi' = 4\pi Q G_N \left(\frac{a}{k}\right)^2 (\rho + p)\theta - {\mathcal H}\psi - \frac{\mathcal{H} \Xi}{a^2}\phi,
\end{equation}

where $\theta$ denotes the divergence of the fluid velocity field. Ref.~\cite{Golovnev:2018wbh} uses the velocity potential $u$, related to $\theta$ through $\theta=-k^2u$.

Combining Eqs.~\eqref{asymm} and \eqref{smix} yields

\begin{equation}\label{eq:grav_slip}
\zeta = \frac{3}{k^2} \left[\left(\frac{\mathcal{H}}{a^2}\left(a^2-\Xi\right)-\frac{\mathcal{H}'}{\mathcal{H}}\right)\phi
+4\pi Q G_N \left(\frac{a}{k}\right)^2(\rho+p)\theta\right].
\end{equation}

This expression explicitly shows that the additional scalar mode is sourced by both the gravitational potential and the velocity perturbations of the matter fluid.

\item The off-diagonal part of the spatial equations:

\begin{equation}\label{eq:pert2}
    \psi = \phi - \mathcal{H}\Xi\,\zeta - 12 \pi Q G_N \left(\frac{a}{k}\right)^2 (\rho+p)\sigma,
\end{equation}

where $\sigma$ denotes the dimensionless scalar anisotropic stress. This equation reveals one of the most characteristic signatures of $f(T)$ gravity. Unlike GR, where a difference between $\psi$ and $\phi$ generally requires anisotropic stress, the extra tetrad mode can generate a gravitational slip even when $\sigma=0$. This effect can become particularly significant on large scales because the constraint for $\zeta$ contains an explicit factor of $k^{-2}$. However, as the background approaches a quasi-de Sitter phase, where $\dot H \equiv (\mathcal{H}'-\mathcal{H}^2)/a^2 \simeq 0$, the quantity $\Xi$ becomes suppressed, provided $Qf_{TT}$ remains finite, and the torsional contribution to the gravitational slip correspondingly diminishes.

\end{itemize}

It is often useful to describe departures from GR through effective phenomenological functions. The perturbation equations \eqref{eq:pert1}, \eqref{smix}, \eqref{eq:pert2}, and \eqref{eq:grav_slip} can therefore be recast as

\begin{eqnarray}
-k^2 \phi &=& 4 \pi G_N a^2 \sum_i \rho_i \Delta_i \, Q,\label{eq:Poisson-like}\\
-k^2 \psi &=& 4 \pi G_N a^2 \sum_i \rho_i \Delta_i \, \mu +12 \pi G_N a^2 \sum_i \rho_i (1+w_i) \sigma_i \, Q,\label{eq:Poisson2-like}\\
-k^2\left(\psi-R \phi\right) &=& 12\pi G_N a^2 \sum_i \rho_i (1+w_i) \left(\sigma_i+\frac{\mathcal{H}\Xi\theta_i}{k^2}\right) \, Q,\label{eq:gslip}\\
-k^2 (\phi+\psi)&=& 8 \pi G_N a^2 \sum_i \rho_i \Delta_i \Sigma + 12 \pi G_N a^2 \sum_i \rho_i (1+w_i) \sigma_i Q,\label{eq:ISW}
\end{eqnarray}
where $\delta_i=\delta\rho_i/\rho_i$, $w_i=p_i/\rho_i$, and
\begin{eqnarray}
     \Delta_i &=& \delta_i + \frac{3\mathcal{H}(1+w_i)}{k^2}\theta_i,\label{eq:fTDelta}\\
     \mu&=& Q\Gamma, \text{ and }\Gamma=\frac{\mu}{Q}=\left[R+\frac{3\mathcal{H}\Xi\sum_i \rho_i (1+w_i) \theta_i}{k^2 \sum_i \rho_i \Delta_i}\right],\label{eq:fTmu}\\
     \Sigma&=&\frac{1}{2}(Q+\mu),\label{eq:fTSigma}\\
     R &=& 1+\frac{3 \Xi}{k^2}\left[\frac{\mathcal{H}^2 \Xi}{a^2}-\mathcal{H}^2+\mathcal{H}'\right].\label{eq:fTR}
\end{eqnarray}

The above parameterization provides a direct connection between theoretical modifications and cosmological observables. In particular, the appearance of gravitational slip in Eq.~\eqref{eq:gslip} demonstrates that departures from GR can occur even without intrinsic anisotropic stress. On sufficiently small scales, where terms proportional to $k^{-2}$ become negligible, $\Gamma=\mu/Q\simeq R$, yielding $\Sigma \simeq Q(1+R)/2$.

The parameters $Q$, $\mu$, and $\Sigma$ are especially valuable because they characterize different observational aspects of gravitational physics. The parameter $Q$ quantifies modifications of the effective gravitational coupling, $\mu$ governs the potential responsible for the motion of non-relativistic matter, while $\Sigma$ controls the lensing potential felt by relativistic particles. Consequently, measurements from galaxy clustering, peculiar velocities, weak lensing, and integrated Sachs-Wolfe (ISW) correlations probe complementary combinations of these quantities.

We note that Eqs.~\eqref{eq:Poisson-like} and \eqref{eq:gslip} reduce to the commonly used parametrized modified-gravity framework of Refs.~\cite{Bertschinger:2008zb,Bean:2010zq}, apart from additional relativistic corrections involving velocity perturbations. The quantity $\mu$ therefore provides a convenient measure of deviations from GR in the growth of cosmic structures, whereas $\Sigma$ characterizes modifications affecting the propagation of light through large-scale structure.

In the GR limit, and in the absence of anisotropic stress, one recovers
\begin{eqnarray}
-k^2 \psi &=& 4 \pi G_N a^2 \sum_i \rho_i \Delta_i,\label{eq:fs_GRPoisson2-like}\\
\psi &=& \phi,\label{eq:fs_GRgslip}
\end{eqnarray}
corresponding to $Q\rightarrow1$, $\mu\rightarrow1$, $\Sigma\rightarrow1$, and $\Xi\rightarrow0$. In this limit, the extra tetrad degree of freedom becomes observationally irrelevant and the standard perturbation equations of GR are fully restored.

For the class of $f(T)$ models considered in this work, deviations from GR are strongly suppressed at early times when the Hubble scale is large. Consequently, significant effects may arise on modes entering the horizon at relatively late stages of the cosmic evolution. These correspond to the largest observable scales, where modifications may become appreciable. Nevertheless, observational detection remains challenging because these scales are also those most strongly affected by cosmic variance.

The diagonal spatial equation contains the pressure perturbation $\delta p$ and is substantially more involved. Owing to the Bianchi identities, together with matter conservation, however, it does not provide independent information and may be derived from the remaining field equations~\cite{Golovnev:2018wbh}.

Tensor perturbations are also straightforward to obtain. For a perfect-fluid matter source their evolution obeys

$$h_{ij}^{\prime\prime}+2{\mathcal H}\left(1 + \frac{\Xi}{2a^2}\right)h_{ij}^{\prime}+k^2 h_{ij}=0,$$

which differs from GR primarily through the modified cosmological background and the corresponding friction term. Importantly, the propagation speed of gravitational waves remains equal to that predicted by GR.

To complete the perturbation framework, the matter sector is described using the perfect-fluid approximation introduced in Eq.~\eqref{eq:matter_Tmn} at the background level, while allowing anisotropic stress $\sigma$ at the perturbative level. Conservation of the perturbed energy-momentum tensor in Fourier space leads to the standard evolution equations~\cite{Ma:1995ey}
\begin{eqnarray}
\delta' &=& -(1+w)\left(\theta-3\phi'\right) + 3{\mathcal H}\left(w - \frac{\delta p}{\delta \rho}\right)\delta\,,\\
\theta' &=& -{\mathcal H}\left(1-3w\right)\theta - \frac{w'}{1+w}\theta + \frac{\delta p/\delta \rho}{1+w}k^2 \delta - k^2 \sigma + k^2 \psi.
\end{eqnarray}

These relations describe the evolution of density and velocity perturbations for a single uncoupled fluid. Together with suitable relations governing $\delta p$ and $\sigma$, they complete the set of equations required for the analysis of linear cosmological perturbations in $f(T)$ gravity.

\section{Family of Exponential \texorpdfstring{$f(T)$}{} Theories of Gravity}\label{Sec:expfT_family}

Recent observations hint that the accelerated expansion of the late Universe could be powered by evolving dark energy rather than a cosmological constant $\Lambda$~\cite{DESI:2025zgx}. On the other hand, the severe tension between local measurements of $H_0$ and its value inferred from Planck assuming flat $\Lambda$CDM motivates dark energy models beyond $\Lambda$CDM~\cite{Planck:2018vyg,CosmoVerseNetwork:2025alb}. We consider a family of $f(T)$ teleparallel gravity theories
\begin{equation}\label{eq:exp_fT}
f(T)=(T+\alpha\sqrt{T_0T})e^{\beta (T_0/T)^b},
\end{equation}
where $T_0=6H_0^2$ in the ($-,+,+,+$) metric signature, while $\alpha$, $\beta$, and $b$ are dimensionless parameters, with $b>0$. Following the procedure given in~\cite{Awad:2017yod}, by evaluating the Friedmann equation at the present time, we obtain a relation between the model parameters. Notably, only two of these parameters are independent at fixed $\Omega_\text{T}$ in the most general case:
\begin{equation}
\beta =
\begin{cases}
\ln{(1-\Omega_\text{T})} &, \text{if } \alpha = -1, \\[5pt]
\frac{1}{2b(1+\alpha)} + W_k\left(-\frac{(1-\Omega_\text{T})e^{-\frac{1}{2b(1+\alpha)}}}{2b(1+\alpha)}\right),&\, \text{otherwise.}
\end{cases}\label{eq:model_parameter}
\end{equation}
The notation $W_k$ denotes the Lambert-$W$ function, where $k=0$ denotes the principal branch and $k=-1$ denotes the secondary real branch, when it exists. Equivalently, for $\beta\ne0$, we may write
\[\alpha=-\frac{1-\Omega_\text{T}-e^\beta+2 b \beta e^\beta}{2b \beta e^\beta}.\]
The particular case $\alpha=0$ and $b=1$, referred to as the exponential infrared $f(T)$ model, has been investigated in~\cite{Awad:2017yod,Hashim:2020sez,Hashim:2021pkq,Akarsu:2024nas,Hashim:2026yoy}. 
Another particular case, $\alpha=-1$, eliminates the dependence on $b$ from the present-day normalization condition, although $b$ remains in the cosmological dynamics. Fixing $b=1$ then reduces the cosmological parameter space to six parameters, as in $\Lambda$CDM. This particular case is investigated in a separate study~\cite{Bedair:2026zvn}. Depending on the model parameter values, teleparallel gravity can mimic evolving dark energy in the phantom regime, a quintessence-to-phantom transition, or a negative-to-positive transition of the torsional (dark energy) density~\cite{Awad:2017yod,Hashim:2020sez,Hashim:2021pkq,Santos:2022atq}. Since the effective dark energy sector is sourced by geometric effects rather than a physical fluid, these behaviors do not require introducing a fundamental phantom fluid; however, stability must be assessed separately. In the present work, we confront the $\alpha=0$ EExpIR model with cosmological data, assessing its ability to accommodate local measurements of the Hubble constant $H_0$ and to fit the combined datasets.

\subsection{The Extended Exponential Infrared (EExpIR) \texorpdfstring{$f(T)$}{} model}

\begin{figure}[h]
    \centering
    \begin{subfigure}[b]{0.52\textwidth}
        \centering
        \includegraphics[width=\textwidth]{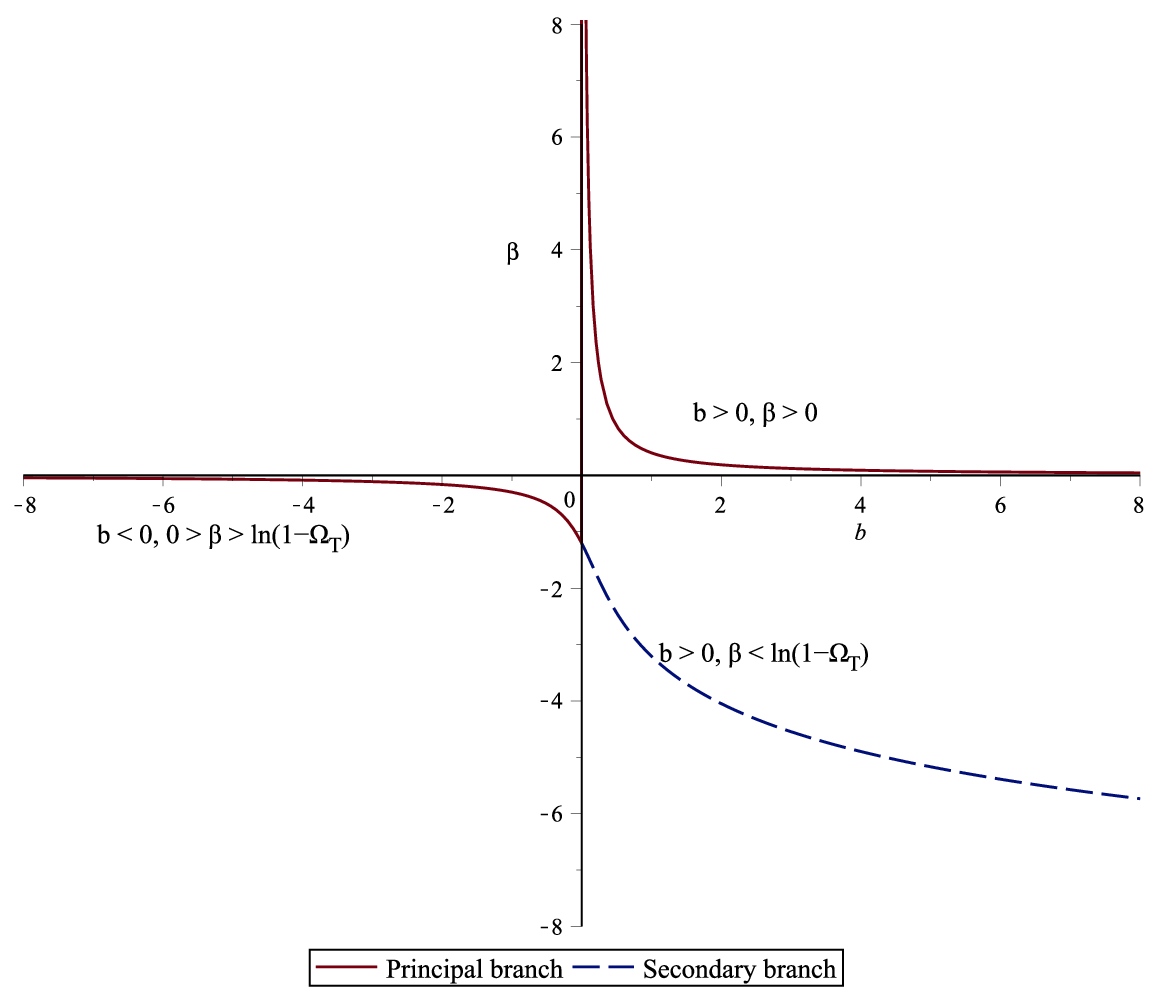}
        \caption{The solution branches.}
        \label{fig:branches}
    \end{subfigure}
   \hfill
    \begin{subfigure}[b]{0.46\textwidth}
    \centering
    \includegraphics[width=\textwidth]{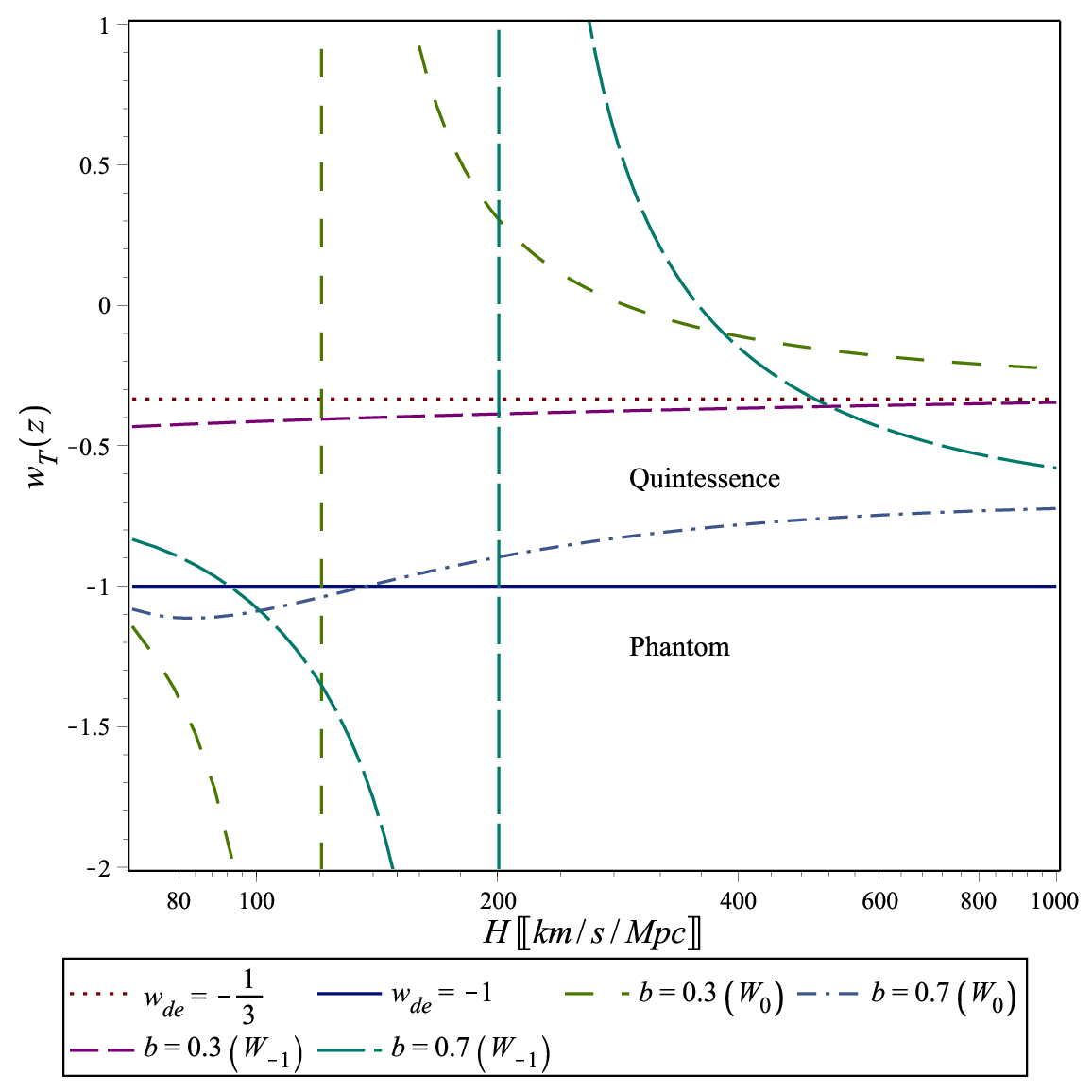}
    \caption{Evolution of the torsion EoS.}
    \label{fig:wde}
    \end{subfigure}
    \caption{(\subref{fig:branches}) The two branches of the Lambert-$W$ solution in Eq.~\eqref{eq:EexpIRfT} for $b\neq0$, excluding the $\alpha=-1$ case. The principal branch is split into two intervals: $b>0$, where $\beta>0$, and $b<0$, where $0>\beta>\ln{(1-\Omega_\text{T})}$. The secondary branch provides a real solution for all $b>0$, where $\beta<\ln{(1-\Omega_\text{T})}$. (\subref{fig:wde}) The evolution of the torsion EoS, Eq.~\eqref{eq:EexpIR_TEoS}, using the two branches, the principal $W_0$ and the secondary $W_{-1}$, for $b=0.3$ and $b=0.7$, representative of $0<b<1/2$ and $1/2<b<1$, respectively.}\label{fig:wde_branches}
\end{figure}

In the present work, we choose the particular case $\alpha=0$ in Eq.~\eqref{eq:exp_fT}, giving
\begin{equation}\label{eq:EexpIRfT}
f(T)=Te^{\beta (T_0/T)^b}, \text{ where }\beta=\frac{1}{2b}+W_k\left(\frac{\Omega_\text{T}-1}{2be^{\frac{1}{2b}}}\right) \text{ and } b\neq 0.
\end{equation}
The extended exponential infrared $f(T)$ teleparallel gravity above generalizes the $\alpha=0$ and $b=1$ case~\cite{Awad:2017yod,Hashim:2020sez,Hashim:2021pkq,Akarsu:2024nas,Hashim:2026yoy} by introducing an additional parameter, $b$. The above theory has been investigated in Ref.~\cite{Santos:2022atq}, where background and structure-growth data were considered. In the present work, we test the theory with CMB data in addition to late-Universe observations. The theory introduces two parameters, $\beta$ and $b$, only one of which is independent. In Fig.~\ref{fig:wde_branches}(\subref{fig:branches}), we plot the two branches of the Lambert-$W$ function in the ($b,\beta$) plane. The case $b=0$ is excluded from this Lambert-$W$ representation. For $0<\Omega_\text{T}<1$, the principal branch is split into two intervals, $b<0$ and $b>0$, the former being associated with $\beta \in (\ln{(1-\Omega_\text{T})},0)$ and the latter with $\beta>0$. We note that $\beta\to \infty$ as $b\to 0^+$ and $\beta\to \ln{(1-\Omega_\text{T})}$ as $b\to 0^-$. On the principal branch, $\beta\to0$ as $|b|\to\infty$, recovering the GR background evolution at finite redshift. The secondary branch provides a real solution for all $b>0$, associated with $\beta<\ln{(1-\Omega_\text{T})}$.

For $b<0$, the exponential term can be expanded as $e^{\beta (T_0/T)^b} \sim 1+\beta (T/T_0)^n+\cdots$, where $n=-b>0$, provided $|\beta(T/T_0)^n|\ll1$. The modification becomes significant at large $T$ (early Universe), while in the low-$T$ limit the cosmic evolution approaches the standard cold dark matter (sCDM) evolution without accelerated expansion. In this sense, we exclude $b<0$ cases since we focus on late-time evolution. In the following subsection, we analyze the phase portraits of the two branches for different values of $b>0$, identifying the viable possibilities.

In order to obtain the cosmic evolution, we substitute \eqref{eq:EexpIRfT} into \eqref{eq:T_dens} and \eqref{eq:T_press}, using Eq.~\eqref{eq:Hdot} with $w=0$. Thus, we evaluate the corresponding effective torsion density and pressure
\begin{eqnarray}
    \rho_T&=& \frac{3H^2}{\kappa^2}\left[1- \left(1- 2\beta b \left(\frac{H_0}{H}\right)^{2b} \right)e^{\beta \left(\frac{H_0}{H}\right)^{2b}}\right],\label{eq:EexpIR_Tdens}\\
    p_T&=&-\frac{3 H^2 \, \beta b \left(\frac{H_0^2}{H^2}\right)^b \left(-\frac{1}{2} + b + \beta b \left(\frac{H_0^2}{H^2}\right)^b
\right)}{\kappa^2 \left[\frac{1}{2}+ \beta^2 b^2 \left(\frac{H_0^2}{H^2}\right)^{2b}+ \beta b \left(b - \frac{3}{2}\right)\left(\frac{H_0^2}{H^2}\right)^b
\right]}.\label{eq:EexpIR_Tpress}
\end{eqnarray}
For $\beta=0$, the torsion density and pressure vanish, $\rho_T=0$ and $p_T=0$, and then the GR limit is recovered. Also, the above equations show clearly that the deviation from GR becomes significant in the low-Hubble regime in the late Universe. In addition, we write the corresponding torsion equation-of-state parameter \eqref{eq:T_EoS}
\begin{equation}\label{eq:EexpIR_TEoS}
    w_T=\frac{
b\beta \left(\frac{H_0}{H}\right)^{2b}
\left(1 - 2b - 2b\beta \left(\frac{H_0}{H}\right)^{2b}\right)
}{
\left[
1- e^{\beta \left(\frac{H_0}{H}\right)^{2b}}
+ 2b\beta \left(\frac{H_0}{H}\right)^{2b}
e^{\beta \left(\frac{H_0}{H}\right)^{2b}}
\right]
\left[
1 - 3\beta b \left(\frac{H_0}{H}\right)^{2b}
+ 2\beta b^2 \left(\frac{H_0}{H}\right)^{2b}
+ 2\beta^2 b^2 \left(\frac{H_0}{H}\right)^{4b}
\right]
}.
\end{equation}
We investigate the evolution of the torsion EoS for different values of the model parameter in the intervals $0<b<1/2$ and $1/2<b<1$ using the two branches in Eq.~\eqref{eq:EexpIRfT}. We visualize these cases in Fig.~\ref{fig:wde_branches}(\subref{fig:wde}). The EoS diverges at a finite Hubble value on the principal branch for $0<b<1/2$ and on the secondary branch for $1/2<b<1$. This feature is associated with a transition from negative to positive dark energy density. On the other hand, the principal branch with $1/2<b<1$ shows a smooth transition from quintessence to phantom dark energy, while for the secondary branch with $0<b<1/2$ the torsional EoS evolves within the quintessence regime. However, at large Hubble values the EoS asymptotically behaves as $w_T\to-b$ in all these cases.

\subsection{Cosmological Evolution and Phase Portrait Analysis}

As mentioned before, we exclude the case where $b<0$, which is relevant to the early Universe rather than late-time evolution. However, the case $b>0$ produces different cosmic evolutions according to the chosen branch, see Fig.~\ref{fig:wde_branches}(\subref{fig:branches}), and the value of the model parameter $b$. In our investigation, in this subsection, we utilize the $f(T)$ phase portrait analysis~\cite{Awad:2017yod} to select the more reasonable choices. Substituting \eqref{eq:EexpIRfT} into \eqref{eq:fT_FR1} and \eqref{eq:fT_FR2}, we write the matter density and pressure
\begin{eqnarray}
    \rho&=&\frac{3H^2}{\kappa^2} \, e^{\beta \left(\frac{H_0}{H}\right)^{2b}}
\left(1 - 2b \beta \left(\frac{H_0}{H}\right)^{2b} \right),\label{eq:EexpIR_Mdens}\\
    p&=&-\frac{4e^{\beta \left(\frac{H_0}{H}\right)^{2b}} \,\dot{H}}{\kappa^2} \left[\frac{1}{2} + b\beta \left(b-\frac{3}{2}\right) \left(\frac{H_0}{H}\right)^{2b}+ \beta^2 b^2 \left(\frac{H_0}{H}\right)^{4b}\right]- \rho.\label{eq:EexpIR_Mpress}
\end{eqnarray}
It proves convenient to use the dimensionless Hubble parameter $E=H/H_0$. Then, we write the distortion function \eqref{eq:distortion} corresponding to the $f(T)$ gravity \eqref{eq:EexpIRfT} as
\begin{equation}
    y=\frac{E^2 \left[1 + e^{\beta E^{-2b}}\left(2\beta b E^{-2b} - 1\right)\right]}{\Omega_T}.
\end{equation}
Substituting $y$ into Eq.~\eqref{eq:E_FR}, we write the corresponding Hubble evolution
\begin{equation}\label{eq:EexpIR_Hubble}
    E^2 \left(1 - 2 b\beta E^{-2b}\right) e^{\beta E^{-2b}}=\Omega_m (1+z)^3+\Omega_r (1+z)^4.
\end{equation}
Clearly, for $\beta=0$, the standard GR version of the Friedmann equation is recovered. However, $\Lambda$CDM with a nonzero cosmological constant cannot be recovered by setting the model parameters $b$ and $\beta$ to particular values.

We define the density parameter of an $i$th component as $\Omega_i=\rho_i/\rho_c$, where $\rho_c=\frac{3}{\kappa^2}H^2$ denotes the critical density. Recalling Eqs.~\eqref{eq:EexpIR_Mdens} and \eqref{eq:EexpIR_Tdens} and neglecting radiation at late times, we obtain the matter and torsion density parameters
\begin{eqnarray}
    \Omega_m(E)&=& e^{\beta E^{-2b}}\left(1 -2 b \beta E^{-2b} \right),\label{eq:EexpIR_MD_para}\\
    \Omega_T(E)&=&1 + \left(2b \beta E^{-2b} - 1\right) e^{\beta E^{-2b}}.\label{eq:EexpIR_TD_para}
\end{eqnarray}

\begin{figure}
    \centering
    \begin{subfigure}[b]{0.48\textwidth}
        \centering
        \includegraphics[width=\textwidth]{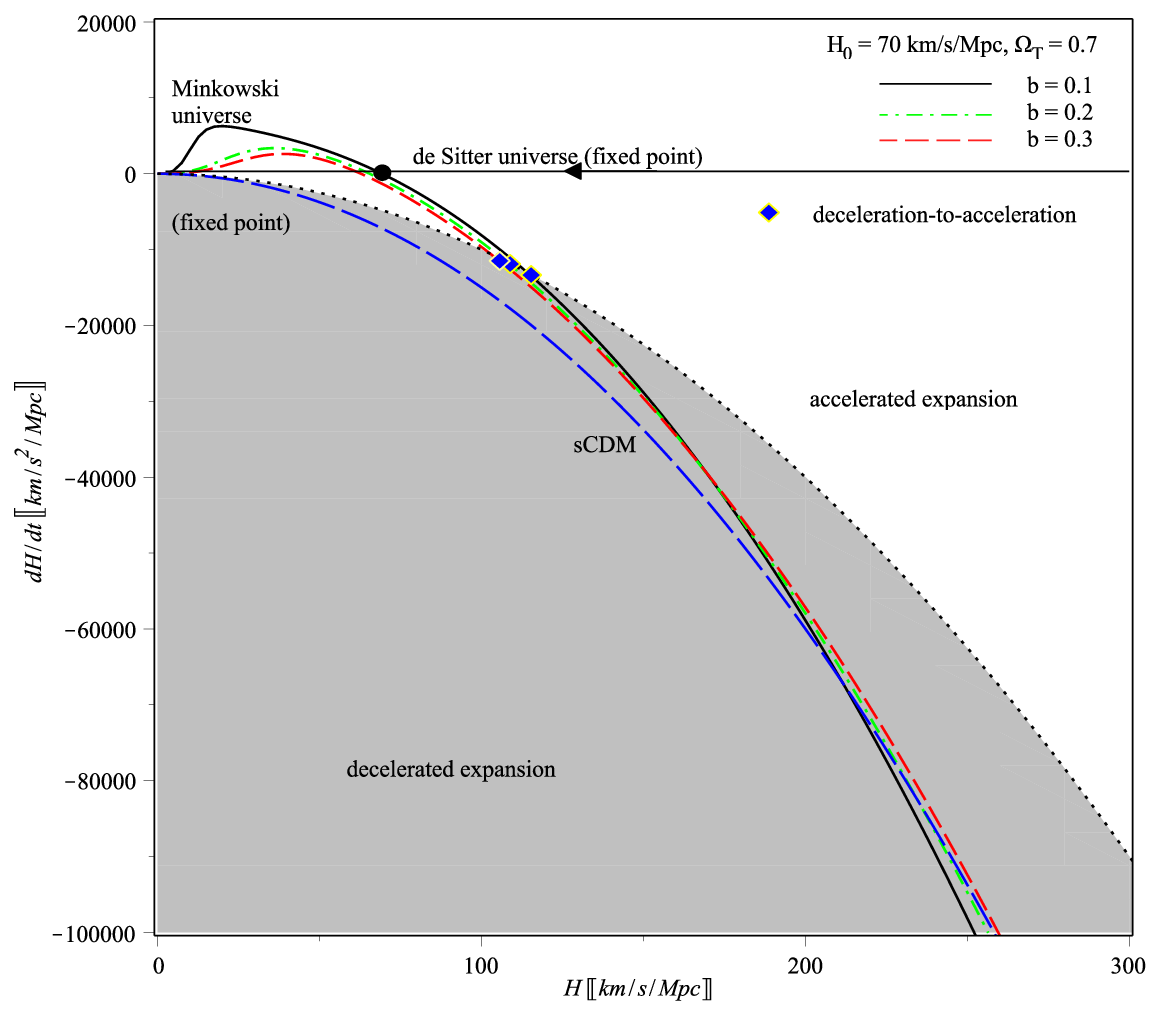}
        \caption{Primary branch ($b<0.5$)}
        \label{fig:ph_port_P1}
    \end{subfigure}
    \hfill
    \begin{subfigure}[b]{0.48\textwidth}
        \centering
        \includegraphics[width=\textwidth]{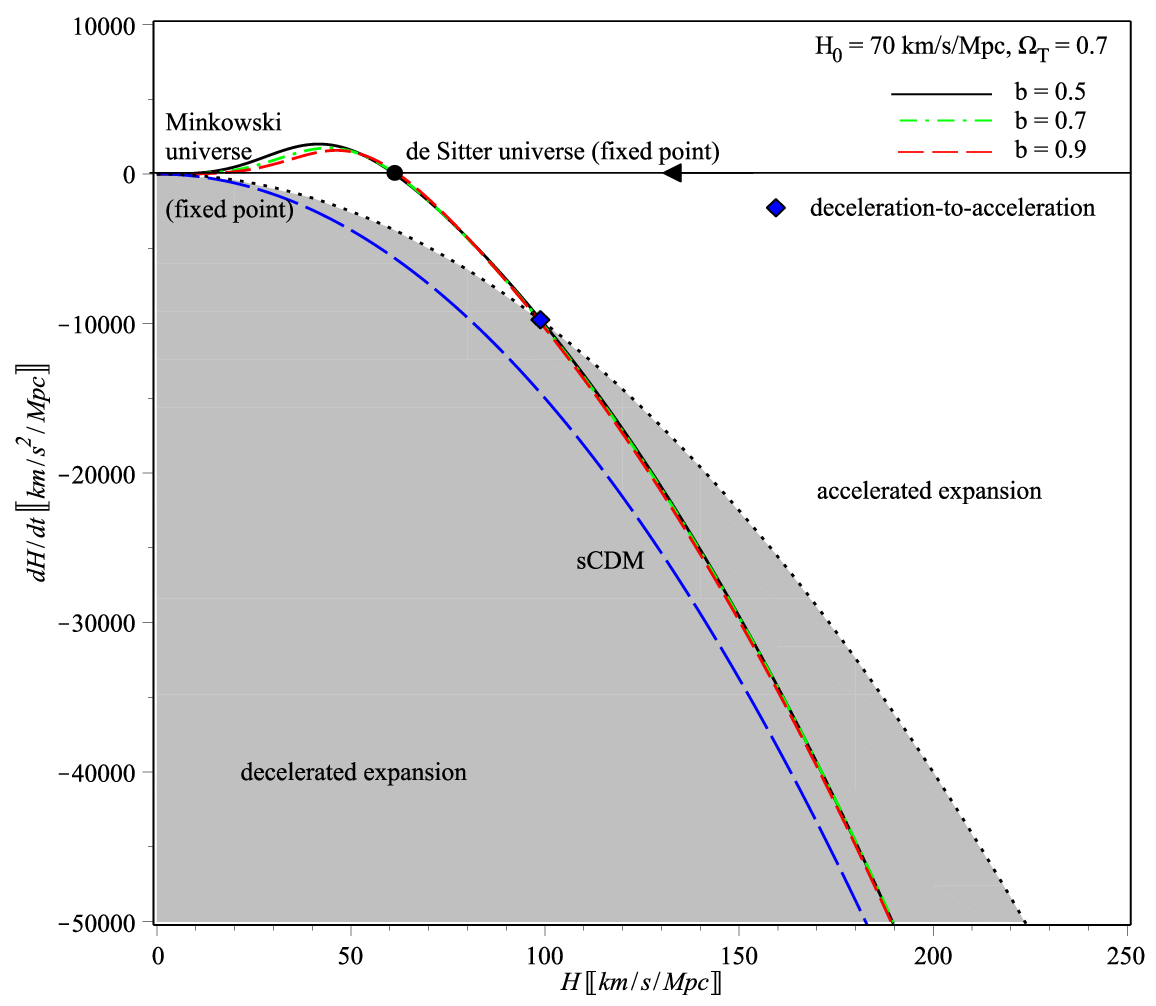}
        \caption{Primary branch ($b>0.5$)}
        \label{fig:ph_port_P2}
    \end{subfigure}
    \hfill
    \begin{subfigure}[b]{0.48\textwidth}
    \centering
    \includegraphics[width=\textwidth]{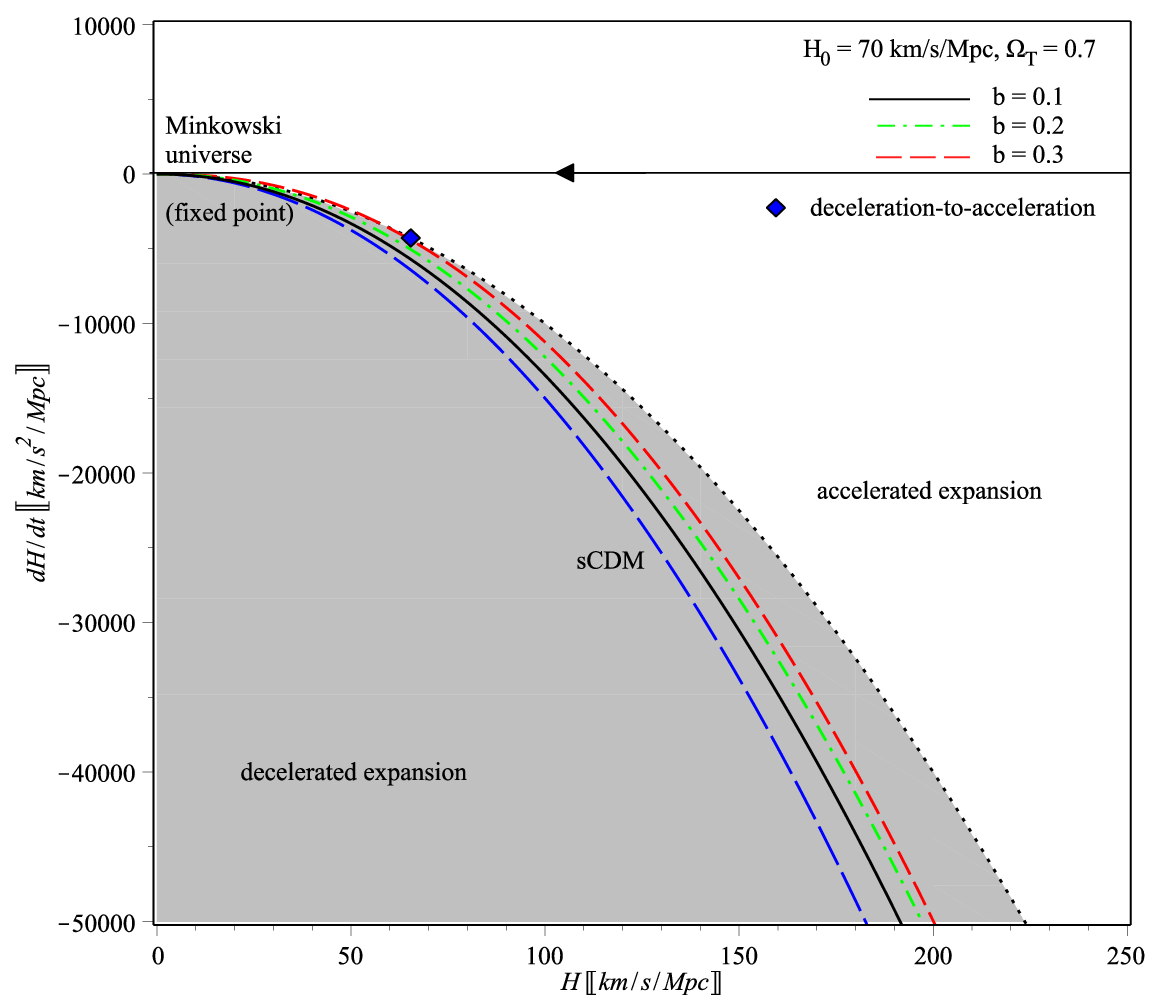}
    \caption{Secondary branch ($b<0.5$)}
    \label{fig:ph_port_S1}
    \end{subfigure}
    \hfill
    \begin{subfigure}[b]{0.48\textwidth}
    \centering
    \includegraphics[width=\textwidth]{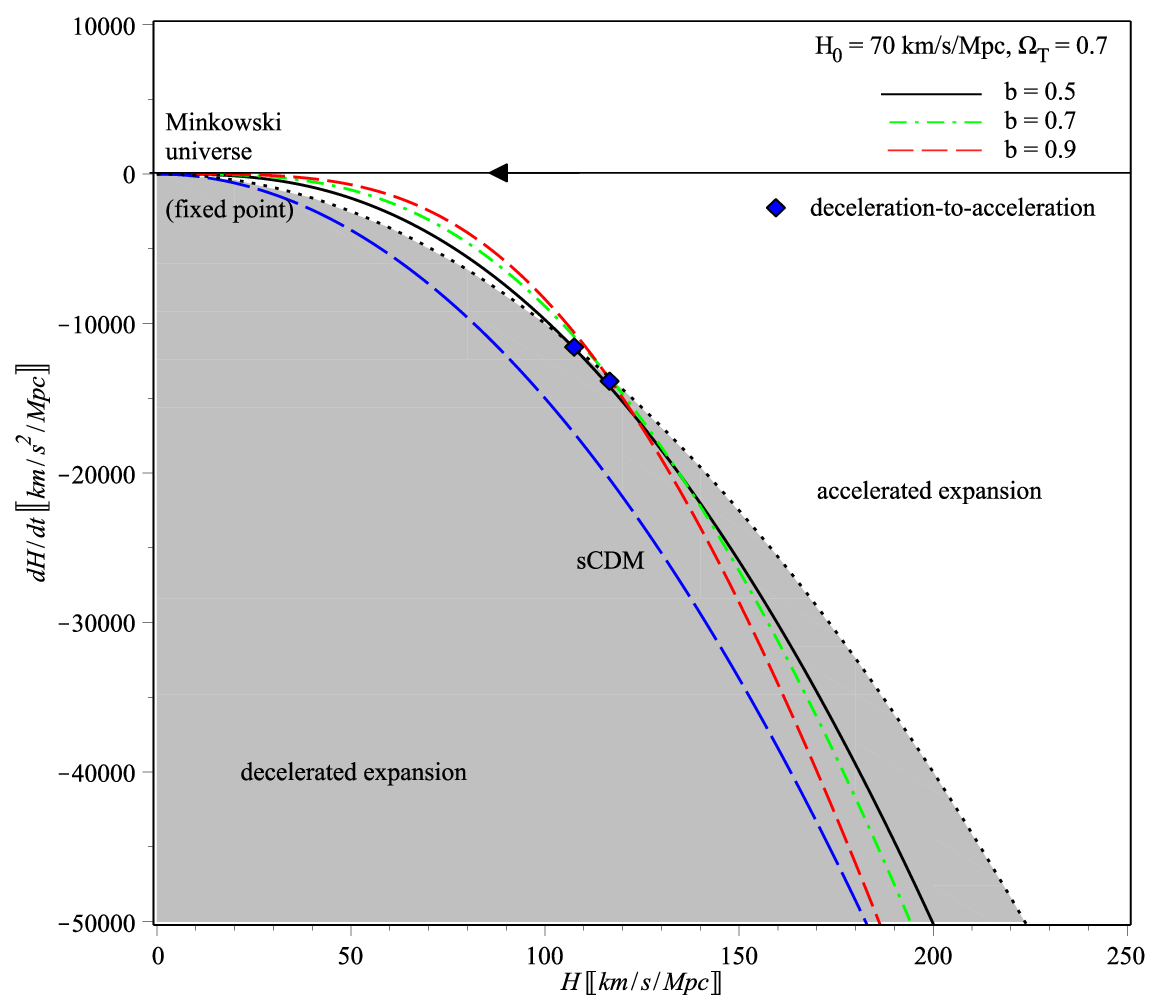}
    \caption{Secondary branch ($b>0.5$)}
    \label{fig:ph_port_S2}
    \end{subfigure}
    \caption{Phase portrait analysis of the two branches of the Lambert-$W$ function in Eq.~\eqref{eq:EexpIRfT}, excluding $\alpha=-1$ and $b<0$. (\subref{fig:ph_port_P1}) The principal branch with $0<b<0.5$: the Universe evolves toward a de Sitter universe at late times, but it modifies the sCDM era (we exclude this case). (\subref{fig:ph_port_P2}) The principal branch with $0.5<b<1$: the Universe evolves toward a de Sitter universe at late times, while preserving the early-time sCDM evolution (we consider this case). (\subref{fig:ph_port_S1}) The secondary branch with $0<b<0.5$: the Universe evolves toward a Minkowski universe at late times, without or with a very weak accelerated expansion phase (we exclude this case). (\subref{fig:ph_port_S2}) The secondary branch with $0.5<b<1$: the Universe evolves toward a Minkowski universe with an accelerated expansion phase at late times, but it modifies the sCDM era (we exclude this case).}\label{fig:phase_portrait}
\end{figure}

It has been shown that the governing dynamical equation corresponding to $f(T)$ cosmology can be written as a one-dimensional autonomous system where $\dot{H}=F(H)$~\cite{Awad:2017yod}. This can be considered a natural extension of GR, which admits the same autonomous description~\cite{Awad:2013tha}. The phase portrait of the $f(T)$ gravity \eqref{eq:EexpIRfT} can be obtained via \eqref{eq:Hdot}
\begin{equation}\label{eq:EexpIR_Hdot}
    \dot{H}=-\frac{3 (1+w) H^2 \left[\frac{1}{2} - \beta b \left(\frac{H_0}{H}\right)^{2b} \right]}{1 + 2b\beta\left(b - \frac{3}{2}\right)\left(\frac{H_0}{H}\right)^{2b} + 2 b^2 \beta^2 \left(\frac{H_0}{H}\right)^{4b}}.
\end{equation}

For $H\to\infty$ and $b>0$, the asymptotic behaviour is $\dot{H}\sim-\frac{3}{2}(1+w)H^2$, matching the quadratic phase portrait of GR cosmology. Then, we expect deviations from GR to be suppressed in the early Universe. This is expected, since $f(T)/T\to1$ as $T\to\infty$. Recalling the previous discussion about the role of the parameter $b$ in obtaining different cosmological evolutions, we omit the case $b<0$, which is more relevant to early-Universe modifications than late-Universe evolution. In addition, it has been shown that the case $b=1$ on the principal branch produces purely phantom-like effective dark energy~\cite{Awad:2017yod}. Similarly, on this branch, $b>1$ cases produce more strongly phantom-like behaviour and may be disfavored by observations, as they can worsen the fit to BAO distances if the early Universe is kept fixed to the standard model~\cite{El-Zant:2018bsc,Hashim:2020sez,Hashim:2021pkq}. Therefore, we omit the case $b\geq1$ in our analysis. Moreover, $b=1/2$ is a critical value, as noted in~\cite{Santos:2022atq}. In this sense, we consider the intervals $0<b<0.5$ and $0.5<b<1$ separately in the phase portrait analysis.

In Fig.~\ref{fig:phase_portrait}, we plot the phase portrait \eqref{eq:EexpIR_Hdot} for four different cases, classified as follows. We consider two cases of the principal branch, i.e., the Lambert-$W_{0}$ function in Eq.~\eqref{eq:EexpIRfT}:
\begin{itemize}
    \item [(a)] The first case is $0<b<0.5$, Fig.~\ref{fig:phase_portrait}(\subref{fig:ph_port_P1}). The phase portrait begins at a finite-time singularity (big bang), with $\dot{H}\to-\infty$ as $H\to\infty$. Also, it crosses the zero-acceleration curve (dotted line), i.e., $\dot{H}=-H^2$, which shows the capability of the model to address late accelerated expansion. The phase portrait evolves toward a de Sitter fixed point, $\dot{H}=0$. Notably, the phase portrait does not match sCDM at intermediate phases. This is associated with the negative dark energy density produced for $b<0.5$ at some epoch, as given by Eq.~\eqref{eq:EexpIR_TD_para} and noted in Ref.~\cite{Santos:2022atq}.
    \item [(b)] The second case is $0.5<b<1$, Fig.~\ref{fig:phase_portrait}(\subref{fig:ph_port_P2}). The phase portrait begins at a big bang singularity. Also, it crosses the zero-acceleration curve, showing the capability of the model to address late accelerated expansion. The phase portrait evolves toward a de Sitter fixed point. Notably, the phase portrait approximately matches sCDM at intermediate phases, while the effective dark energy can cross from quintessence to phantom at late times~\cite{Santos:2022atq}.
\end{itemize}

Two additional cases correspond to the secondary branch, i.e., the Lambert-$W_{-1}$ function in Eq.~\eqref{eq:EexpIRfT}:
\begin{itemize}
    \item [(c)] The third case is $0<b<0.5$, Fig.~\ref{fig:phase_portrait}(\subref{fig:ph_port_S1}). The phase portrait begins at a big bang singularity and evolves toward a Minkowski fixed point ($H=0$, $\dot{H}=0$). In the intermediate-$H$ regime, the phase portrait deviates from the sCDM phase portrait and produces only a weak accelerated expansion phase over the plotted range.
    \item [(d)] The fourth case is $0.5<b<1$, Fig.~\ref{fig:phase_portrait}(\subref{fig:ph_port_S2}). The phase portrait begins at a big bang singularity and evolves toward a Minkowski fixed point. In the intermediate-$H$ regime, the phase portrait deviates from the sCDM phase portrait, but it can produce a late accelerated expansion phase.
\end{itemize}

The phase portraits of the four cases above are represented in Fig.~\ref{fig:phase_portrait}. We exclude cases (a), (c), and (d), where the model does not approximately reproduce the sCDM phase portrait in the intermediate-Hubble regime. Thus, our choice is to work with case (b), i.e., the principal branch where $0.5<b<1$, according to the phase portrait analysis. This choice is also motivated by the results previously obtained for the $f(T)$ gravity \eqref{eq:EexpIRfT} in Ref.~\cite{Santos:2022atq}, where background and structure-growth analyses give central estimates spanning $b\sim0.67$--$0.87$, depending on the data combination. In the present study, we extend the observational data by considering more recent early-Universe CMB data as well as recent late-Universe data to test the theory. This will be the subject of the following two sections.

\section{Data}\label{Sec:data}

The observational constraints presented in this work are obtained within a Bayesian inference framework implemented through the \texttt{Cobaya} package~\cite{Torrado:2020dgo}\footnote{\href{https://cobaya.readthedocs.io/en/latest/}{https://cobaya.readthedocs.io/en/latest/}.}. Cosmological predictions, including the expansion history, distance measures, and the CMB angular power spectra, are computed using a modified implementation of the \texttt{CLASS} Boltzmann solver~\cite{Blas:2011rf}\footnote{
This code is publicly available at \href{https://github.com/mwhashim/class_tmg}{https://github.com/mwhashim/class\_tmg}.}.

Throughout the analysis we assume a spatially flat cosmological model composed of baryons, radiation, cold dark matter, and an effective dark-energy sector generated by the teleparallel modification of gravity. Relative to the standard $\Lambda$CDM scenario, the considered $f(T)$ model introduces one additional parameter, denoted by $b$, which controls the evolution of the effective torsional dark energy. The parameter vector explored in the Monte Carlo analysis is therefore

\[
\left\{\Omega_\textrm{b} h^2, \Omega_\textrm{c} h^2, H_0, \tau_\textrm{reio},
\log{(10^{10} A_s)}, n_s, b \right\}.
\]

The adopted prior ranges are summarized in Table~\ref{tab:pprior}. Following the standard neutrino treatment employed in contemporary cosmological analyses, we consider a total neutrino mass of $\sum m_\nu = 0.06\,\mathrm{eV}$, implemented as one massive and two effectively massless neutrino species. For the model parameter $b$, we restrict the prior interval to $0.5<b<1.0$, motivated by the phenomenological study of Ref.~\cite{Santos:2022atq}.

\begin{table}[htp]
\centering
 \caption{Adopted priors adopted for the sampled cosmological parameters.}
    \label{tab:pprior}
    \begin{tabular}{|c|c|}
    \Xhline{1.1pt}
    Parameter & Prior\\
    \hline
         $\Omega_\textrm{b} h^2$ & ${\cal U}(0.005, 0.1)$ \\
         $\Omega_\textrm{c} h^2$ & ${\cal U}(0.001, 0.99)$ \\
         $H_0$ & ${\cal U}(40,100)$\\
         $\log(10^{10} A_\mathrm{s})$ & ${\cal U}(1.61, 3.91)$\\
         $n_s$ & ${\cal U}(0.8, 1.2)$\\
         $\tau_\textrm{reio}$ & fixed to 0.06 for purely late universe data, \\
                              & otherwise ${\cal N}(\mu = 0.051, \sigma=0.006)$ \\
         $b$ & ${\cal U}(0.5, 1.0)$\\
         \Xhline{1.1pt}
    \end{tabular}
\end{table}

The posterior distributions are sampled using the Metropolis--Hastings algorithm. To ensure convergence and sampling stability, four independent Markov chains are evolved simultaneously and required to satisfy the Gelman--Rubin convergence criterion $R-1\leq0.02$~\cite{Gelman:1992zz}.

The observational information employed in this work consists of complementary probes of both the background expansion and the growth history of the Universe.

\begin{itemize}

\item \textbf{Cosmic Microwave Background (CMB)}:

The early-Universe constraints are derived from a combined compilation of state-of-the-art CMB measurements obtained from Planck, ACT, and SPT observations. More specifically, the dataset includes

\begin{enumerate}
    \item the Planck 2018 low-multipole temperature likelihood (\texttt{Commander}) covering $2\leq\ell\leq29$~\cite{Planck:2019nip};

    \item the ACT DR6 primary-CMB likelihood based on foreground-marginalized temperature and polarization spectra~\cite{AtacamaCosmologyTelescope:2025blo};

    \item the Planck 2018 high-multipole temperature and polarization likelihood, restricted to $\ell<1000$ for TT and $\ell<600$ for TE and EE, combined with ACT DR6~\cite{Planck:2019nip,AtacamaCosmologyTelescope:2025blo};

    \item the combined ACT DR6 and Planck CMB lensing likelihood~\cite{Carron:2022eyg,ACT:2023dou,ACT:2023kun};

    \item the SPT-3G TT/TE/EE likelihood derived from the 2019--2020 observing campaign~\cite{SPT-3G:2025bzu};

    \item the SPT-3G delensed polarization and lensing analysis based on the 2019--2020 observations and the MUSE methodology~\cite{SPT-3G:2024atg}.
\end{enumerate}

For brevity, this aggregate dataset will be referred to as \textbf{CMB-SPA} throughout the remainder of the paper.

\item \textbf{Baryon Acoustic Oscillations (BAO)}:

To constrain the late-time expansion history, we employ the DESI Data Release 2 BAO measurements (hereafter \textbf{DESI})~\cite{DESI:2025zgx}. These observations provide high-precision determinations of cosmological distances, making them a particularly powerful probe of background cosmology and dark-energy phenomenology.

\item \textbf{Type Ia Supernovae (SNIa)}:

As a complementary distance indicator, we use the PantheonPlus compilation (\textbf{PP})~\cite{Brout:2022vxf}, consisting of 1550 spectroscopically confirmed Type Ia supernovae represented by 1701 light curves obtained from 18 independent surveys. The catalog spans a broad redshift range extending from the local Universe to $z\lesssim2.3$, thereby providing robust constraints on the recent expansion history. When combined with the SH0ES distance-ladder calibration, we denote the resulting dataset as \textbf{PPS}.

\end{itemize}

In order to assess the performance of the model under different observational conditions, we analyze several dataset combinations. Two combinations rely exclusively on late-time measurements, namely PP+DESI and PPS+DESI. Constraints from the early Universe are obtained using the CMB-SPA compilation alone. Finally, joint analyses are performed by combining early- and late-time probes through the datasets CMB-SPA+PP+DESI and CMB-SPA+PPS+DESI. This strategy enables us to evaluate the consistency of the model across different cosmological epochs and to quantify the impact of each observational probe on the inferred parameter constraints.

\section{Results}\label{Sec:results}

\begin{figure}
    \centering
    \includegraphics[width=0.8\linewidth]{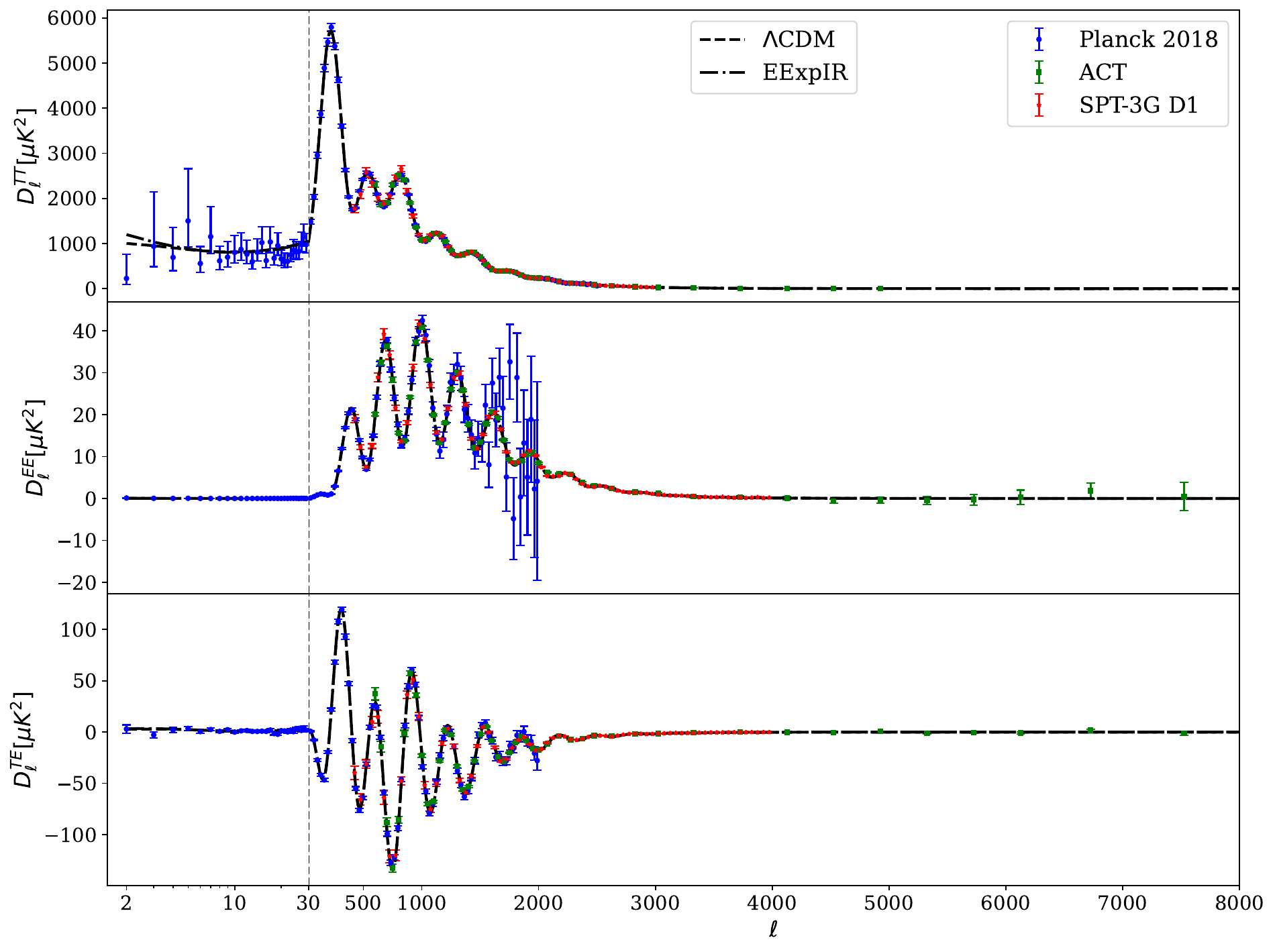}
    \caption{The angular power spectra of $\Lambda$CDM and the $f(T)$ model, where CMB data are from Planck 2018, ACT, and SPT-3G. The dashed line represents $\Lambda$CDM, and the dash-dotted line represents the $f(T)$ model. The latter slightly increases the TT power spectrum at low multipoles, $\ell\lesssim 5$, which can be associated with a modified late-time ISW contribution, including the effects of gravitational slip in $f(T)$ gravity described by Eq.~\eqref{eq:grav_slip}.}
    \label{fig:CMB-power_spectra}
\end{figure}

Before presenting the parameter constraints and statistical analysis, it is instructive to examine the impact of the model on the primary cosmological observables. Figure~\ref{fig:CMB-power_spectra} displays the temperature and polarization angular power spectra (TT, TE, and EE) computed using the best-fit cosmological parameters obtained from the CMB-SPA dataset for both the $\Lambda$CDM and $f(T)$ scenarios, with the corresponding marginalized parameter constraints listed in Table~\ref{tab:model2Tab}.

As expected, the $\Lambda$CDM prediction reproduces the observed CMB anisotropy spectra in excellent agreement with the Planck baseline cosmology~\cite{Planck:2018vyg}. The $f(T)$ model likewise provides a very good description of the combined CMB-SPA observations, yielding angular spectra that remain nearly indistinguishable from those of the standard cosmological model over most multipole ranges. This behavior reflects the fact that the considered teleparallel modification closely approaches the GR limit during the epochs that primarily determine the acoustic structure of the CMB.

Small deviations emerge only on the largest angular scales, where the $f(T)$ model produces a mild enhancement of the TT spectrum at low multipoles ($\ell \lesssim 5$). Physically, this behavior can be associated with changes in the late-time ISW contribution, which depends on the line-of-sight integral of $\phi'+\psi'$. Both the modified background evolution and the scalar perturbation sector discussed in Sec.~\ref{Sec:fT_cosmology} can affect this contribution. In particular, the additional degree of freedom encoded in Eq.~\eqref{eq:grav_slip} contributes to a nonvanishing gravitational slip, modifying the evolution of the metric potentials on very large scales. The explicit $k^{-2}$ dependence in the expression for $\zeta$ allows these corrections to become important on large scales, although their magnitude also depends on the source terms.

Despite this enhancement, the resulting departure from the standard prediction remains modest. As the background approaches a de Sitter phase, $\Xi$ tends to zero, provided that $Qf_{TT}$ remains finite. Consequently, the correction entering Eq.~\eqref{eq:pert2} is suppressed in this limit. Furthermore, the observational significance of these large-scale modifications is weakened by the substantial cosmic variance that affects the lowest CMB multipoles.

Overall, the figure demonstrates that the proposed $f(T)$ cosmology preserves the successful CMB phenomenology of $\Lambda$CDM while allowing for small, theoretically motivated deviations on the largest observable scales. Interestingly, the corresponding goodness-of-fit statistics reported in Table~\ref{tab:model2Tab} indicate a marginally better fit to the CMB-SPA dataset than that obtained within the standard $\Lambda$CDM framework.

\subsection{Cosmological Constraints and Parameter Estimation}

The marginalized constraints obtained for both $\Lambda$CDM and the EExpIR $f(T)$ model are summarized in Table~\ref{tab:model2Tab}. A comparison among the different data combinations reveals distinct responses of the cosmological parameters to late- and early-time observations.

Considering only geometrical probes of the late Universe, namely PP+DESI, both models allow relatively broad ranges for the Hubble constant. In this case, the posterior distribution remains weakly constrained because these datasets alone do not completely break the degeneracies among the background cosmological parameters. The EExpIR model yields a slightly larger central value of $H_0$ than $\Lambda$CDM, but the difference is not statistically significant. Once the SH0ES calibration is incorporated through the PPS+DESI combination, the broad $H_0$ posterior is dominated by the calibration, yielding similarly tight constraints close to $H_0 \simeq 74~{\rm km\,s^{-1}\,Mpc^{-1}}$ in both cosmologies.

\begin{figure}[h]
    \centering
    \begin{subfigure}[b]{0.48\textwidth}
        \centering
        \includegraphics[width=\textwidth]{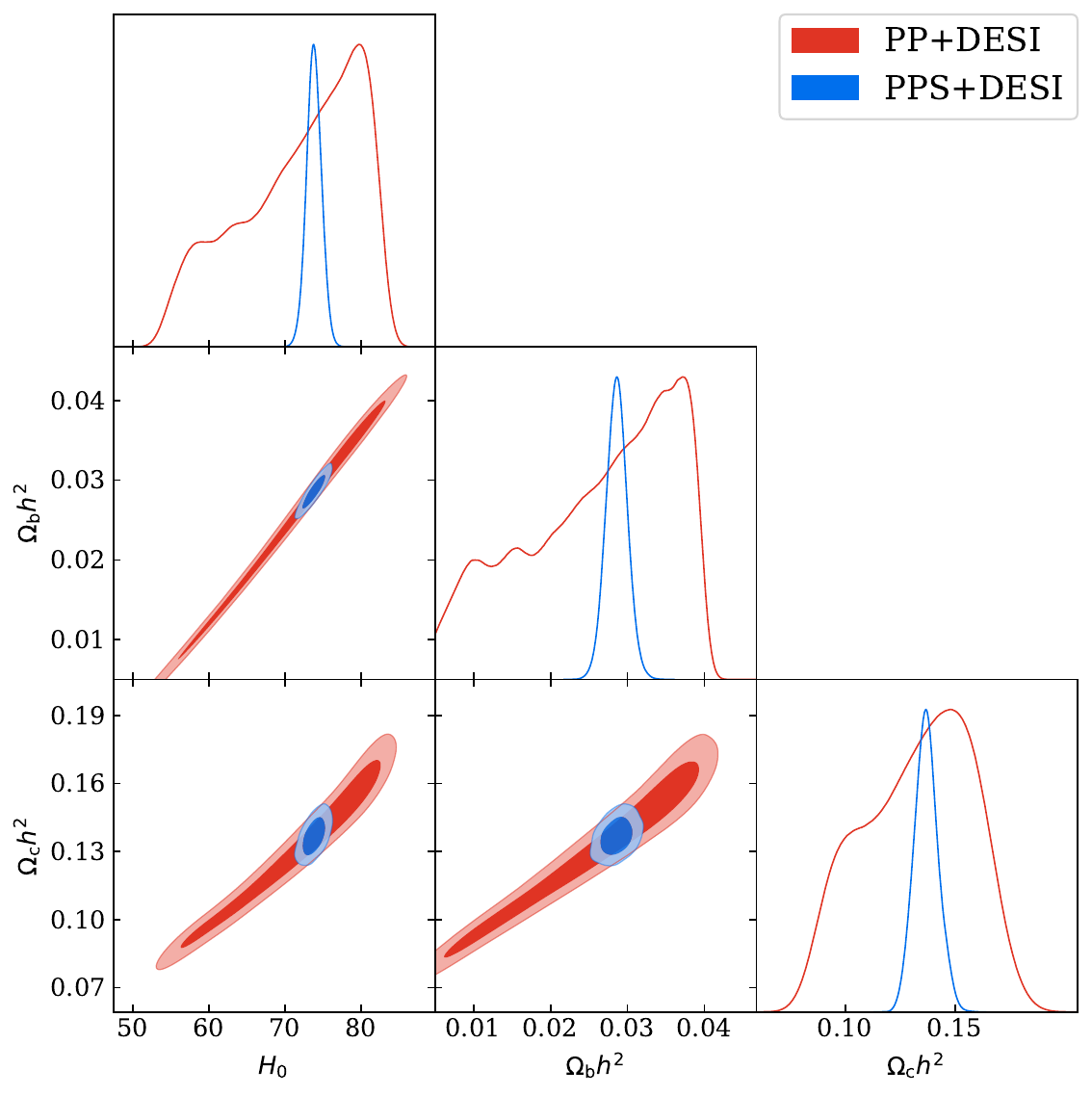}
        \caption{$\Lambda$CDM}
        \label{fig:LCDM_late}
    \end{subfigure}
    \hfill
    \begin{subfigure}[b]{0.48\textwidth}
    \centering
    \includegraphics[width=\textwidth]{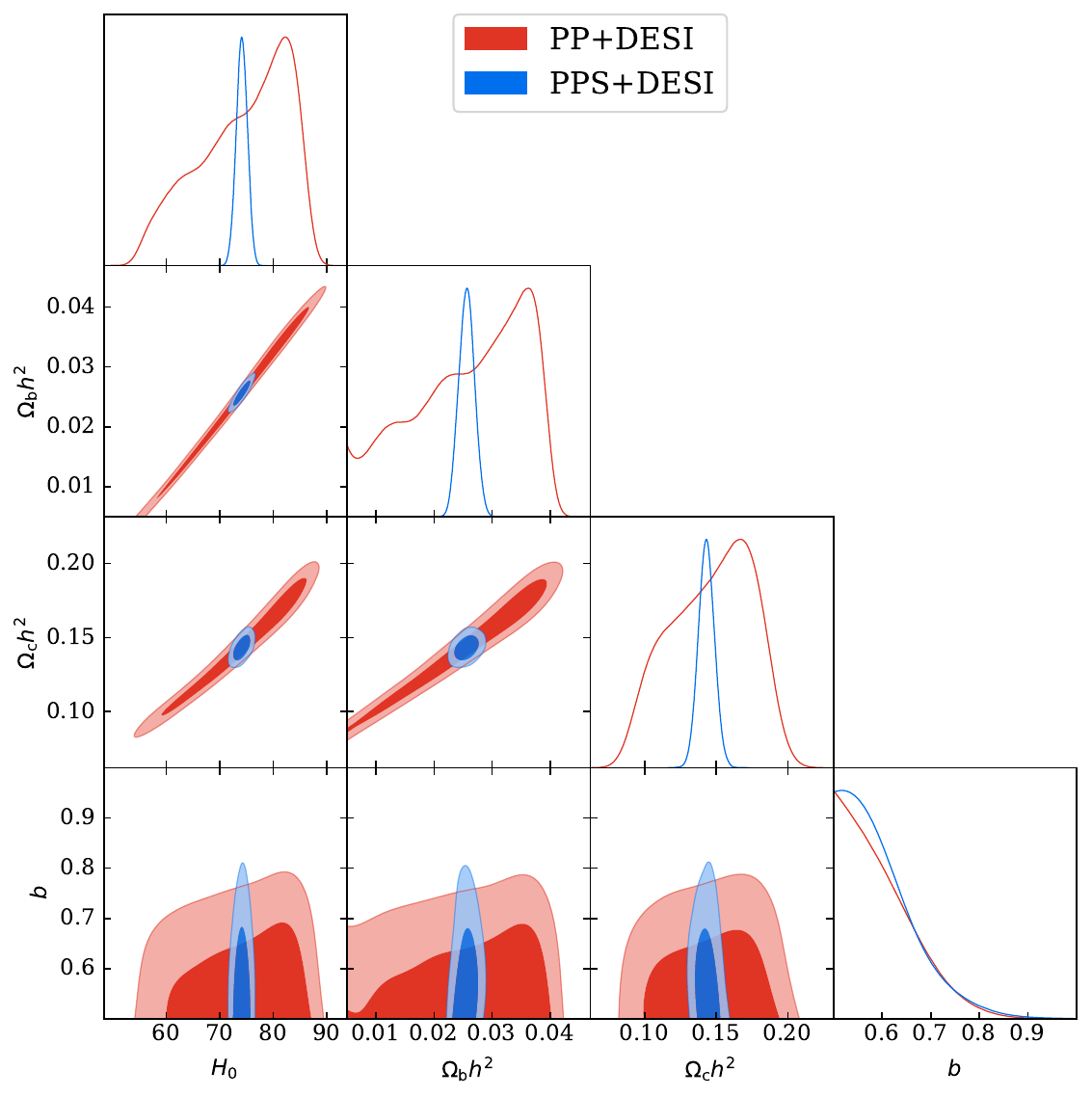}
    \caption{EExpIR}
    \label{fig:fT_late}
    \end{subfigure}
    \caption{Late-Universe constraints from PP+DESI and PPS+DESI data on the $\Lambda$CDM and EExpIR models, showing 68\% and 95\% confidence regions. (\subref{fig:LCDM_late}) For $\Lambda$CDM, the PP+DESI data only weakly constrain $H_0$. Including the SH0ES calibration yields a tighter $H_0$ posterior dominated by the calibration and breaks the geometric degeneracy involving $\Omega_{\rm b} h^2$ and $\Omega_{\rm c} h^2$. (\subref{fig:fT_late}) For EExpIR, similarly to $\Lambda$CDM, the PP+DESI data only weakly constrain $H_0$ and the model parameter $b$. Including SH0ES yields a calibration-dominated $H_0$ posterior and tighter constraints on the other cosmological parameters. Values $b<0.5$ are excluded by the adopted prior, motivated by the phase portrait analysis.} \label{fig:late_data}
\end{figure}

The inferred high-$H_0$ solutions are accompanied by smaller derived values of the sound horizon at decoupling, typically $r_\star \sim 136$--$140~{\rm Mpc}$. This behavior reflects the correlation between $H_0$ and the sound horizon inferred from BAO distances within the assumed cosmological model. The corresponding posterior distributions for $\{\Omega_\mathrm{b}h^2,\Omega_\mathrm{c}h^2,H_0,b\}$ are displayed in Fig.~\ref{fig:late_data}. As expected, the inclusion of SH0ES information substantially improves parameter determination. For the EExpIR model, the preferred region of parameter space obtained from late-time observations corresponds to $b \lesssim 0.62$, allowing a dark-energy evolution capable of crossing the phantom divide within the adopted prior $0.5<b<1$. This result is broadly consistent with the background-level analysis reported in Ref.~\cite{Santos:2022atq}. We next incorporate the CMB-SPA compilation and the full perturbation analysis developed in Sec.~\ref{Sec:fT_cosmology}, allowing a simultaneous assessment of early- and late-time cosmological information.

\begin{table}[h!]
    \centering
    \setlength{\tabcolsep}{10pt}
    \renewcommand{\arraystretch}{1.0}
    \caption{Observational constraints on the model parameters of $\Lambda$CDM and the EExpIR model at 68\% CL. The minimum values of $\chi^2$ for the different datasets are given below, where $\Delta \chi^2=\chi^2_{f}-\chi^2_\Lambda$. The subscript $\star$ denotes the decoupling epoch.}\label{tab:model2Tab}
    \renewcommand{\arraystretch}{1.5}
    \resizebox{\textwidth}{!}{%
    \begin{tabular}{|l|c|c|c|c|c|}
    \Xhline{1.1pt}
    \input{summarytable_LCDM.txt}\\
    \hhline{|=|=|=|=|=|=|}
    \input{summarytable_Model-II.txt}\\
    \Xhline{1.1pt}
    \end{tabular}}
\end{table}

\begin{figure}
    \centering
    \includegraphics[width=\linewidth]{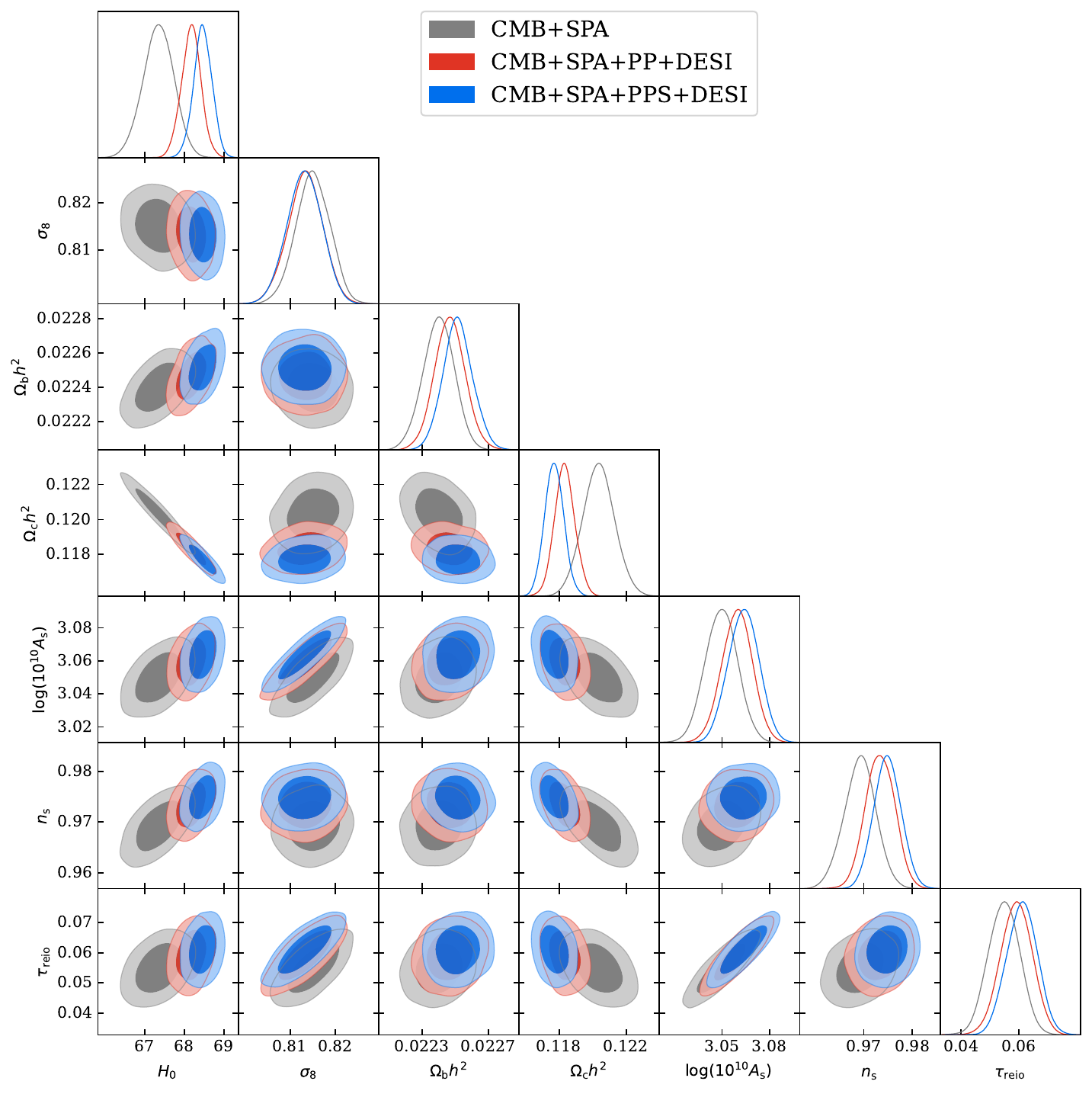}
    \caption{Early- and combined early- and late-Universe constraints on the $\Lambda$CDM model, showing 68\% and 95\% confidence regions. For early-Universe data (CMB-SPA), the cosmological parameters are consistent with the Planck-$\Lambda$CDM results. For the combined datasets (CMB-SPA+PP/PPS+DESI), some of the six base $\Lambda$CDM parameters, including $n_s$ and $\tau_{\rm reio}$, shift toward higher values, generally by amounts of order $1\sigma$ relative to the CMB-SPA-only uncertainties.}
    \label{fig:LCDM_early}
\end{figure}

For the standard $\Lambda$CDM model, the CMB-SPA dataset alone yields parameter values fully consistent with the Planck baseline cosmology~\cite{Planck:2018vyg}. A more interesting picture emerges when late-time observations are added. Combining CMB-SPA with either PP+DESI or PPS+DESI produces modest shifts in several of the six base $\Lambda$CDM parameters. In particular, $\Omega_{\rm b}h^2$, $A_s$, $n_s$, and $\tau_{\rm reio}$ move toward larger values, with shifts generally of order $1\sigma$ relative to the CMB-SPA-only uncertainties, while the derived matter density parameter $\Omega_{\rm m}$ decreases. Since the CMB tightly constrains the physical matter density $\Omega_{\rm m}h^2$, this reduction is naturally associated with the mild increase in the inferred Hubble constant. These shifts reflect the mismatch between the regions of parameter space preferred by CMB measurements and low-redshift observations. Similar trends have also been found in other analyses combining CMB observations with late-time data~\cite{Gorbunov:2026sly,Hashim:2026yoy}. The associated degeneracy directions are illustrated in Fig.~\ref{fig:LCDM_early}. Within $\Lambda$CDM, adding late-time observations shifts the constraints along these degeneracy directions, whereas modified-gravity scenarios may alter the parameter correlations~\cite{Hashim:2026yoy}.

\begin{figure}
    \centering
    \includegraphics[width=\linewidth]{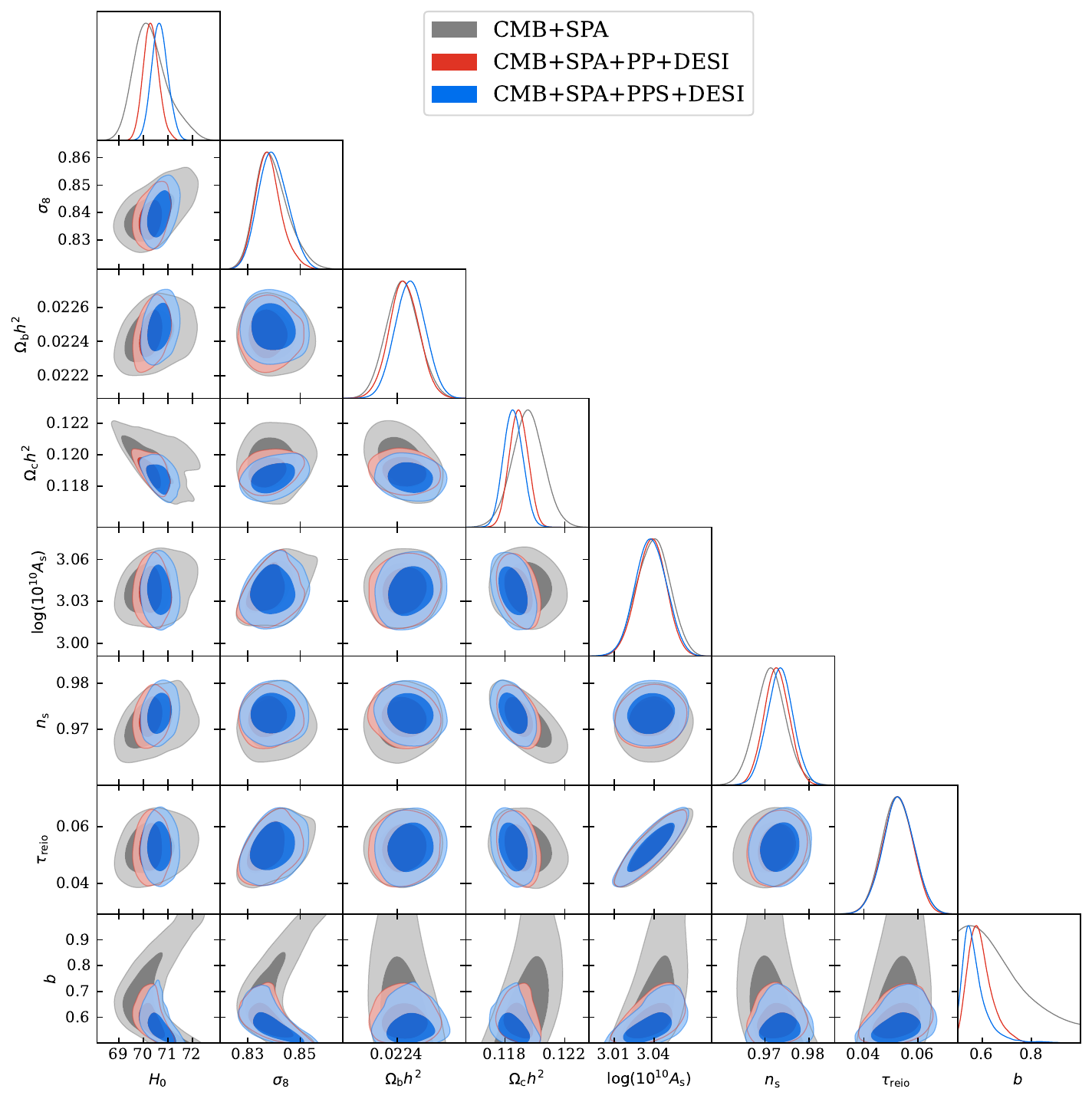}
    \caption{Early- and combined early- and late-Universe constraints on the EExpIR model, showing 68\% and 95\% confidence regions. For early-Universe data (CMB-SPA), the model parameter is weakly constrained, with $b<0.714$ at 68\% CL. For the combined datasets (CMB-SPA+PP/PPS+DESI), most of the six base cosmological parameters show only modest shifts relative to CMB-SPA alone, in contrast to $\Lambda$CDM. The clearest changes are tighter constraints on $H_0$ and $b$ and a shift of $\Omega_{\rm c}h^2$ toward lower values. The full data combination CMB-SPA+PPS+DESI constrains $b$ to $0.513<b<0.589$ at 68\% CL.}
    \label{fig:fT_early}
\end{figure}

For the EExpIR model, the situation is qualitatively different. Using only the CMB-SPA dataset, we obtain

\[
H_0=70.32^{+0.52}_{-0.82}\ {\rm km\,s^{-1}\,Mpc^{-1}},
\]

which already lies closer to local determinations than the corresponding $\Lambda$CDM value. Although the CMB-SPA dataset alone does not tightly constrain the extra model parameter $b$, the addition of late-Universe observations leads to a remarkably stable parameter reconstruction. As seen in Fig.~\ref{fig:fT_early}, the six base cosmological parameters exhibit only modest shifts when PP+DESI or PPS+DESI data are included. This stability contrasts with the behavior observed in $\Lambda$CDM, but reflects the rigidity of the model rather than a better fit to the observations. While the inferred $H_0$ is closer to local determinations, the low-redshift expansion history lacks sufficient flexibility to accommodate the combined data, resulting in a poorer fit than $\Lambda$CDM.

The combined datasets constrain the EExpIR parameter to $b=0.592^{+0.031}_{-0.059}$ for CMB-SPA+PP+DESI and $b=0.573^{+0.016}_{-0.060}$ for CMB-SPA+PPS+DESI at 68\% CL. These constraints are considerably tighter than those obtained previously from background observables alone~\cite{Santos:2022atq}, highlighting the constraining power of the combined background and perturbation analysis with high-precision CMB measurements. However, this tighter determination does not resolve the model's difficulty in fitting the low-redshift observations.

\begin{figure}
    \centering
    \includegraphics[width=0.5\linewidth]{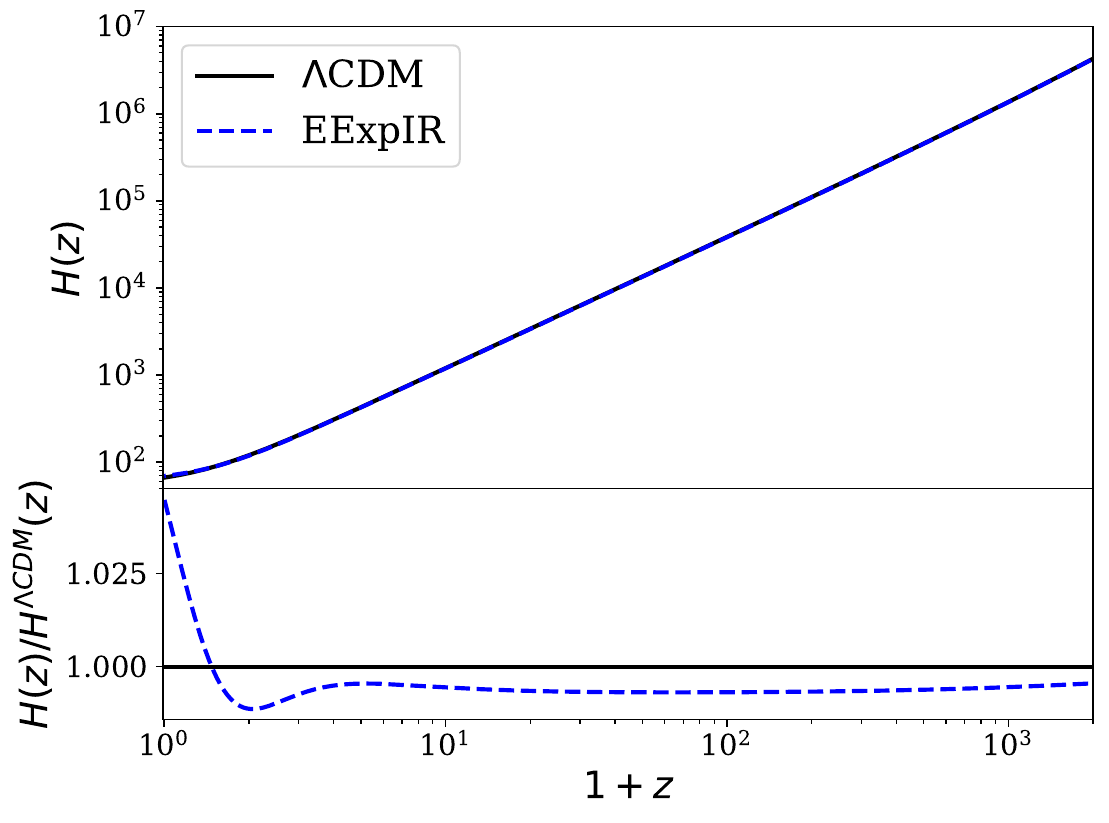}
    \caption{The upper panel shows the Hubble evolution for the $\Lambda$CDM and EExpIR models, using the best-fit parameters of the CMB-SPA+PPS+DESI dataset. The EExpIR evolution is given by Eq.~\eqref{eq:EexpIR_Hubble}. The corresponding parameter constraints are listed in Table~\ref{tab:model2Tab}. The lower panel shows the ratio of the EExpIR Hubble expansion rate to that of $\Lambda$CDM. The EExpIR expansion rate is slightly lower at intermediate and high redshifts, but becomes higher near the present epoch, yielding a larger $H_0$. This behavior accompanies the transition of the effective torsional dark energy from quintessence to phantom-like behavior.}
    \label{fig:Hubble_evolution}
\end{figure}

To further illustrate the impact of the modified gravity sector, Fig.~\ref{fig:Hubble_evolution} presents the Hubble expansion history predicted by Eq.~\eqref{eq:EexpIR_Hubble} using the best-fit parameters of the CMB-SPA+PPS+DESI dataset. The figure shows that the EExpIR model remains close to $\Lambda$CDM throughout most of the cosmic evolution while predicting a moderately larger present-day expansion rate. This feature illustrates how the model shifts $H_0$ toward local determinations without introducing large departures from the standard early-Universe expansion history.

The lower panel of Fig.~\ref{fig:Hubble_evolution} shows that the relative difference with respect to $\Lambda$CDM remains small at high redshift. The EExpIR expansion rate is slightly lower at intermediate and high redshifts, but becomes higher near the present epoch. Consequently, the largest deviations occur in the recent Universe, where the torsional contribution acts as an effective dynamical dark-energy component. Despite the higher $H_0$, the rigidity of the model's low-redshift expansion history prevents it from accommodating the combined observations as well as $\Lambda$CDM. This difficulty is consistent with the restrictions that BAO and supernova distances place on late-time solutions to the Hubble tension~\cite{Lemos:2018smw}.

Additional insight into the model is provided by the effective torsional equation of state. According to Eq.~\eqref{eq:EexpIR_TEoS}, the asymptotic early-time behavior in the matter-dominated approximation approaches $w_T \rightarrow -b$. Given the preferred range of $b$, the torsional sector initially behaves as a quintessence-like component. As the Universe evolves and infrared corrections become increasingly relevant, the equation of state gradually crosses the cosmological-constant boundary and enters a phantom-like regime.

\begin{figure}[h]
    \centering
    \begin{subfigure}[b]{0.45\textwidth}
    \centering
    \includegraphics[width=\textwidth]{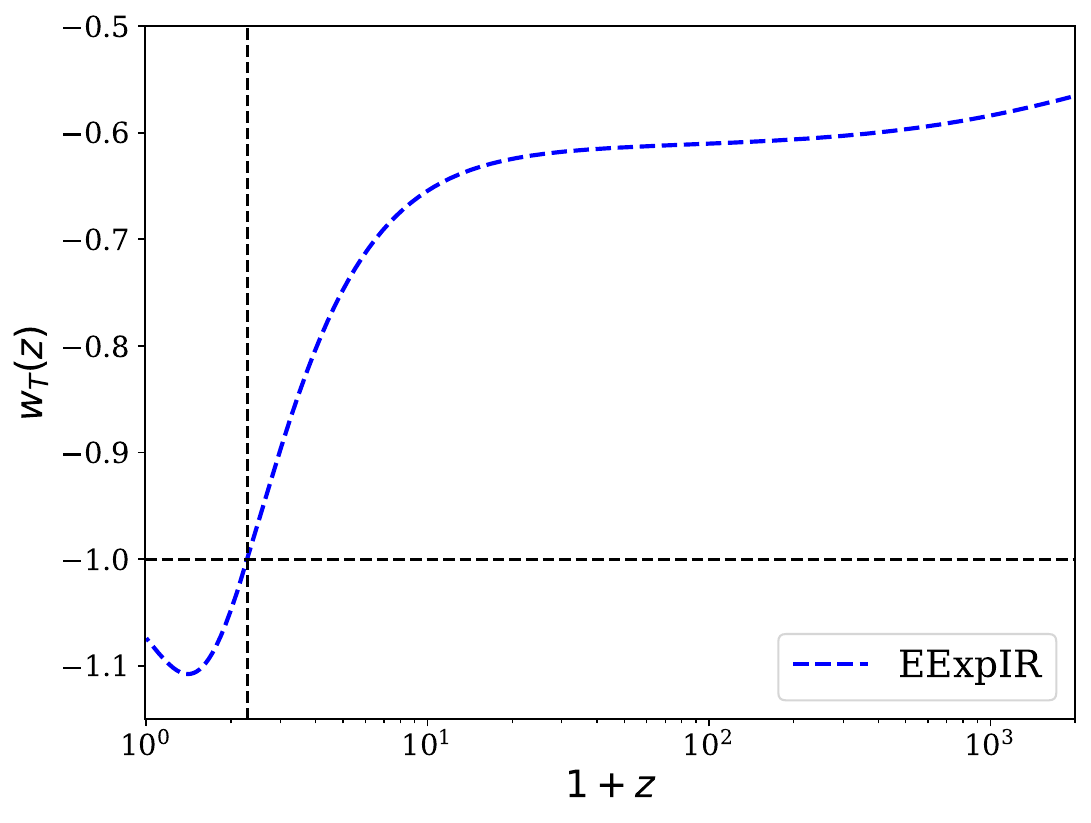}
    \caption{Torsion EoS}
    \label{fig:TEoS}
    \end{subfigure}
    \hfill
     \begin{subfigure}[b]{0.45\textwidth}
        \centering
        \includegraphics[width=\textwidth]{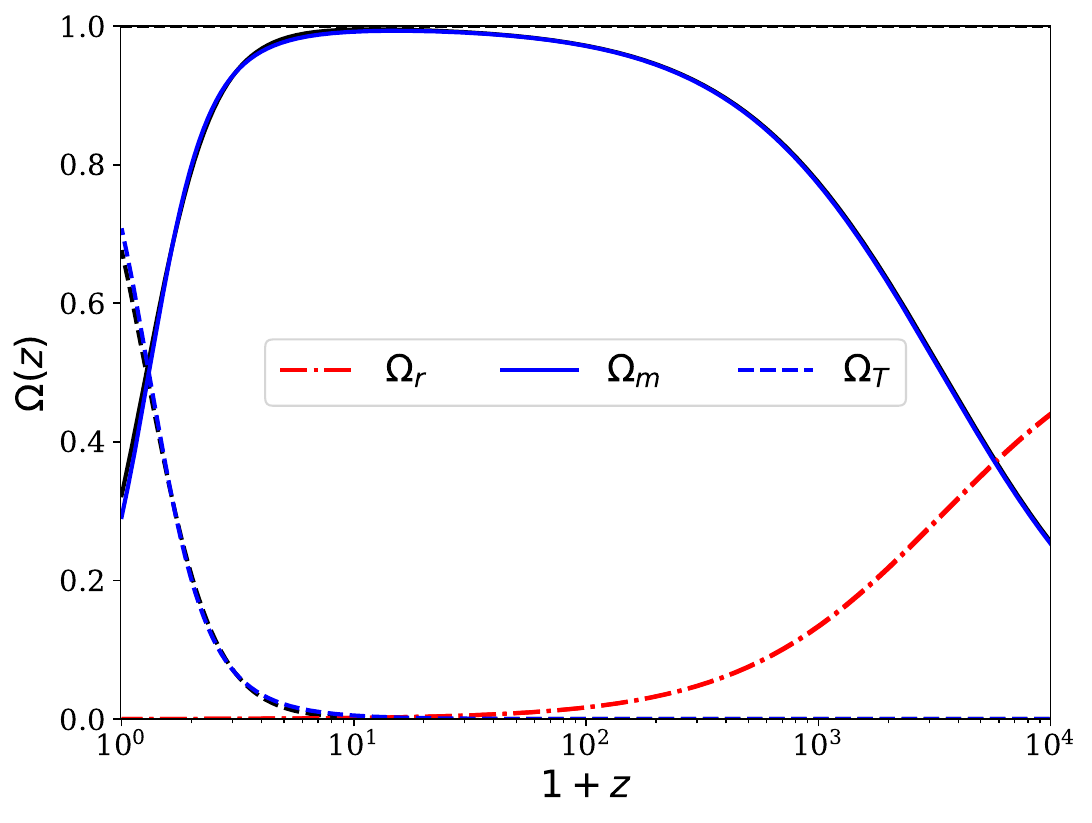}
        \caption{Density parameters}
        \label{fig:dens_param}
    \end{subfigure}
    \caption{Cosmic evolution of the EExpIR model. (\subref{fig:TEoS}) The torsional EoS evolution, Eq.~\eqref{eq:EexpIR_TEoS}, where modified gravity mimics dynamical dark energy with a smooth transition from quintessence to phantom at redshift $z_{\rm trans}=1.29$. In the early-time matter-dominated approximation, $w_T\to-b$. (\subref{fig:dens_param}) The evolution of the density parameters, with the matter and torsional components given by Eqs.~\eqref{eq:EexpIR_MD_para} and \eqref{eq:EexpIR_TD_para}, respectively. Torsional dark energy becomes dominant at late times, qualitatively as in the $\Lambda$CDM scenario.}\label{fig:Torsion_evolution}
\end{figure}

The transition point may be estimated by imposing $w_T=-1$, which yields

\begin{equation}
    H_{\rm trans}=H_0\left(\frac{\beta}{x_{\rm trans}}\right)^{\frac{1}{2b}},
\end{equation}

where $x_{\rm trans}$ is the nonzero root of
$e^{x_{\rm trans}}\left[1+b(2b-3)x_{\rm trans}+2b^2x_{\rm trans}^2\right]=1$
corresponding to the phantom-divide crossing, and $\beta$ is given by Eq.~\eqref{eq:EexpIRfT}. The resulting evolution is displayed in Fig.~\ref{fig:Torsion_evolution}(\subref{fig:TEoS}), indicating a transition near $z\simeq1.29$. The corresponding evolution of the matter and torsional density parameters, shown in Fig.~\ref{fig:Torsion_evolution}(\subref{fig:dens_param}), confirms that the torsional contribution gradually overtakes the matter sector at late times, reproducing the accelerated expansion of the Universe in a manner compatible with the standard cosmological picture.

Overall, the cosmological analysis highlights both the advantages and limitations of the EExpIR model. For CMB-SPA alone, the model provides a slightly improved fit relative to $\Lambda$CDM, with $\Delta\chi^2=-2.24$, while simultaneously yielding a larger value of the Hubble constant. When all probes are combined, the standard model achieves the lower total $\chi^2$, corresponding to $\Delta\chi^2=7.27$. For CMB-SPA+PPS+DESI, the EExpIR model yields

\[
H_0=70.67^{+0.31}_{-0.35}\ {\rm km\,s^{-1}\,Mpc^{-1}},
\]

which remains closer to local determinations than the corresponding $\Lambda$CDM value. However, the stability of the inferred parameters across the CMB-SPA and combined datasets reflects the rigidity of the model's low-redshift expansion history, which prevents it from fitting the combined observations as well as $\Lambda$CDM. The balance between a higher $H_0$ and a poorer combined fit motivates a more detailed examination of statistical consistency measures and model-comparison metrics, which we present in the following subsection.

\subsection{\texorpdfstring{$H_0$}{} Tension and Model Selection}

An important benchmark for any extension of the standard cosmological model is its ability to accommodate the discrepancy between early- and late-Universe determinations of the Hubble constant. The so-called $H_0$ tension originates from the difference between the value inferred from CMB observations within the $\Lambda$CDM framework and that obtained from direct local measurements. Since modified gravity models alter the expansion history and the growth of perturbations, they provide a possible avenue for reducing this disagreement.

To quantify the level of disagreement between the inferred values of $H_0$ and the local determinations from SH0ES~\cite{Riess:2021jrx} and H0DN~\cite{H0DN:2025lyy}, we employ the Gaussian tension estimator~\citep{DES:2020hen}

\begin{equation}
    N_{\sigma} = \frac{|\mu_D - \mu_{\rm obs}|}{\sqrt{\sigma_D^2 + \sigma^2_{\rm obs}}},
\end{equation}

where $\mu_D$ and $\sigma_D$ denote the mean value and uncertainty inferred from a given dataset combination, while $\mu_{\rm obs}$ and $\sigma_{\rm obs}$ correspond to the local measurement under consideration. This estimator assumes independent, approximately Gaussian constraints. The resulting estimates are summarized in Table~\ref{tab:H0_Tensions}. Comparisons between SH0ES/H0DN and combinations including PPS are omitted, since the SH0ES calibration is already included in PPS and these comparisons are therefore not independent tension tests.

\begin{table}
\centering
\caption{Gaussian tension between the $H_0$ values inferred from different dataset combinations and the local measurements from SH0ES ($H_0=73.04\pm1.04~{\rm km\,s^{-1}\,Mpc^{-1}}$~\cite{Riess:2021jrx}) and H0DN ($H_0=73.50\pm0.81~{\rm km\,s^{-1}\,Mpc^{-1}}$~\cite{H0DN:2025lyy}). Combinations including PPS are omitted because they already incorporate the SH0ES calibration, which also contributes to H0DN, making these comparisons non-independent.}
\label{tab:H0_Tensions}
\begin{tabular}{|l|c|c|c|}
\Xhline{1.1pt}
\multirow{2}{*}{Model} & \multirow{2}{*}{Dataset} & \multicolumn{2}{c|}{Gaussian Tension with $H_0$} \\ \cline{3-4}
 & & \multicolumn{1}{c|}{~~~~SH0ES~~~~} & \multicolumn{1}{c|}{H0DN} \\ \hline
\multicolumn{1}{|c|}{\multirow{3}{*}{$\Lambda$CDM}} & PP+DESI & \multicolumn{1}{c|}{$0.10\sigma$} & \multicolumn{1}{c|}{$0.15\sigma$} \\ \cline{2-4}
\multicolumn{1}{|c|}{} & CMB-SPA & \multicolumn{1}{c|}{$5.13\sigma$} & $6.85\sigma$ \\ \cline{2-4}
\multicolumn{1}{|c|}{} & CMB-SPA+PP+DESI & \multicolumn{1}{c|}{$4.55\sigma$} & $6.29\sigma$ \\ \hline
\multirow{3}{*}{EExpIR} & PP+DESI & \multicolumn{1}{c|}{$0.19\sigma$} & \multicolumn{1}{c|}{$0.15\sigma$} \\ \cline{2-4}
 & CMB-SPA & \multicolumn{1}{c|}{$2.34\sigma$} & $3.30\sigma$ \\ \cline{2-4}
 & CMB-SPA+PP+DESI & \multicolumn{1}{c|}{$2.53\sigma$} & $3.71\sigma$ \\
\Xhline{1.1pt}
\end{tabular}
\end{table}

Several interesting trends emerge from Table~\ref{tab:H0_Tensions}. For $\Lambda$CDM, the CMB-SPA dataset leads to a substantial discrepancy with local measurements, corresponding to tensions of $5.13\sigma$ and $6.85\sigma$ relative to SH0ES and H0DN, respectively. Even after incorporating PP+DESI, the disagreement remains significant. This behavior reflects the well-known tendency of the standard cosmological model to favor a lower value of the present-day expansion rate when constrained primarily by early-Universe information.

The EExpIR model exhibits a markedly different behavior. Using the same CMB-SPA dataset, the tension is reduced to $2.34\sigma$ relative to SH0ES and $3.30\sigma$ relative to H0DN. The discrepancy is also smaller than in $\Lambda$CDM when PP+DESI is included. For the CMB-SPA+PPS+DESI combination, the Gaussian estimator decreases from $4.29$ to $2.18$ with respect to SH0ES and from $5.97$ to $3.26$ with respect to H0DN. However, these latter values describe residual offsets rather than independent tension significances, since the SH0ES calibration is included in PPS and also contributes to H0DN. The independent comparisons using CMB-SPA and CMB-SPA+PP+DESI therefore show that EExpIR reduces, but does not eliminate, the discrepancy between cosmological and local determinations of the Hubble constant.

A natural question is whether this improvement is achieved at the expense of introducing excessive model complexity. To address this issue, we complement the tension analysis with a statistical comparison between EExpIR and the reference $\Lambda$CDM cosmology.

As a first diagnostic, we employ the Akaike Information Criterion (AIC)~\cite{Akaike:1974vps,Liddle:2007fy},

\begin{equation}
    \label{eqn:aic}
    \text{AIC}_\text{model} \equiv -2\ln \mathcal{L}_\text{max} +2k
    = \chi^2_\text{min} +2k,
\end{equation}

where $\mathcal{L}_\text{max}$ denotes the maximum likelihood, $k$ is the number of free parameters, and $\chi^2_\text{min}$ is the minimum chi-square value. The AIC balances goodness of fit against model complexity through an explicit penalty on the number of parameters. We define
\[
\Delta\text{AIC} \equiv \text{AIC}_\text{model}-\text{AIC}_{\Lambda\mathrm{CDM}},
\]
such that positive values favor the reference $\Lambda$CDM scenario according to the AIC.

Since the AIC uses only the maximum likelihood and the number of free parameters, we additionally compute the Bayesian evidence
\begin{equation}
    \mathcal{Z}_i = \mathcal{Z}(d \, | \, \mathcal{M}_i )
    = \int d\theta \,
    p(d \, | \, \theta,\mathcal{M}_i)\,
    p(\theta \, | \, \mathcal{M}_i),
\end{equation}
using \texttt{Harmonic}\footnote{\href{https://astro-informatics.github.io/harmonic/index.html}{https://astro-informatics.github.io/harmonic/index.html}}~\citep{Polanska:2024arc}. Here, $p(d|\theta,\mathcal{M}_i)$ denotes the likelihood function and $p(\theta|\mathcal{M}_i)$ is the prior probability density. The quantities $d$, $\theta$, and $\mathcal{M}_i$ represent the observational data, model parameters, and cosmological model, respectively.

The Bayesian comparison between competing models is performed through the Bayes factor
\begin{equation}\label{eqn:bayesfac}
    \ln\mathcal{B}_{ij}
    \equiv
    \ln\mathcal{Z}_i
    -
    \ln\mathcal{Z}_j,
\end{equation}
where model $j$ corresponds to the reference $\Lambda$CDM cosmology. Following the Jeffreys scale~\cite{Trotta:2008qt}, values satisfying $|\ln\mathcal{B}_{ij}|<1$ are considered statistically inconclusive, while larger values indicate increasing levels of preference for one model over the other.\footnote{The interval $1\lesssim |\ln\mathcal{B}_{ij}| \lesssim 2.5$ indicates weak evidence for one model over the other, the interval $2.5\lesssim |\ln\mathcal{B}_{ij}| \lesssim 5$ indicates a moderate preference, and $|\ln\mathcal{B}_{ij}| \gtrsim 5$ corresponds to strong evidence.} Within the adopted convention, negative values imply a preference for $\Lambda$CDM.

While Bayesian evidence accounts for both goodness of fit and prior-volume effects, the AIC remains useful as a simple and transparent indicator of predictive performance. For this reason, both statistics are considered in the present analysis.

\begin{table}
    \centering
     \caption{Model-selection statistics for $\Lambda$CDM and EExpIR across the dataset combinations considered. Differences in $\chi^2$ and AIC are defined as the EExpIR values minus the corresponding $\Lambda$CDM values; positive $\Delta\mathrm{AIC}$ favors $\Lambda$CDM. The log Bayes factor is defined as $\ln\mathcal{B}_{ij}=\ln\mathcal{Z}_{\rm EExpIR}-\ln\mathcal{Z}_{\Lambda\mathrm{CDM}}$, with negative values favoring $\Lambda$CDM. The strength of Bayesian evidence is classified using the adopted Jeffreys scale.}
    \begin{tabular}{l|l|c|c|c|c|c|c}
        \Xhline{1.1pt}
        Model
            & Dataset & $\chi^2$ & $\Delta\chi^2$ & AIC & $\Delta$AIC & $\ln\mathcal{B}_{ij}$ & Strength of Bayesian evidence \\
        \hline
           \multirow{5}{*}{$\Lambda$CDM}
            & PP+DESI & 1416.17  & 0.0 & 1426.168  &  0.0 &-- &-- \\
            & CMB+SPA & 604.17  & 0.0 &  616.17  & 0.0 & -- &--\\
            & CMB+SPA+PP+DESI & 2024.51  & 0.0 & 2036.506  & 0.0 & -- &--\\
            & PPS+DESI & 1465.35  & 0.0 & 1475.352 & 0.0 & -- &-- \\
            & CMB+SPA+PPS+DESI & 2103.85  & 0.0 & 2115.848 & 0.0 & -- &-- \\
        \hhline{:========}
            \multirow{5}{*}{EExpIR}
            & PP+DESI & 1433.44  & 17.27  & 1445.438  & 19.27 & $-$\,\,8.66 & ~~Strong preference of $\Lambda$CDM\\
            & CMB+SPA & 601.92  & $-$2.245 & 615.925  & $-$0.245 &  +\,\,0.45& ~~~Statistically indistinguishable\\
            & CMB+SPA+PP+DESI & 2047.58  & 23.074 & 2061.58  & 25.075 & $-$9.38  & ~~Strong preference of $\Lambda$CDM\\
            & PPS+DESI & 1482.7  & 17.348  & 1494.7  & 19.348 & $-$\,\,8.37 & ~~Strong preference of $\Lambda$CDM\\
            & CMB+SPA+PPS+DESI & 2111.12  & 7.27 & 2125.118  & 9.27 & $-$\,\,1.61 & Weak preference of $\Lambda$CDM\\
        \Xhline{1.1pt}
    \end{tabular}
    \label{tab:BE_harmonics}
\end{table}

The resulting model-selection statistics are presented in Table~\ref{tab:BE_harmonics}. For the CMB-SPA dataset alone, the EExpIR model achieves a slightly lower $\chi^2$ than $\Lambda$CDM, corresponding to $\Delta\chi^2=-2.24$, while the Bayes factor remains close to zero, $\ln\mathcal{B}_{ij}=0.45$. According to the Jeffreys criterion, the comparison is therefore inconclusive, consistent with the negligible AIC difference. Although EExpIR also yields a higher $H_0$, the CMB-only constraints retain correlations and degeneracies that allow different late-time expansion histories to reproduce similar CMB spectra. Low-redshift observations are therefore essential to break these degeneracies and test whether the expansion history associated with the higher $H_0$ is consistent with the data.

The situation changes when additional low-redshift information is incorporated. BAO and supernova distances constrain the late-time expansion history, exposing the rigidity of the EExpIR model: its additional parameter does not provide sufficient freedom to retain the higher $H_0$ while reproducing the distance measurements as well as $\Lambda$CDM. For the full dataset combination CMB-SPA+PPS+DESI, the Bayes factor, $\ln\mathcal{B}_{ij}=-1.61$, indicates weak evidence for $\Lambda$CDM, while the AIC also favors the standard model. The inclusion of the SH0ES calibration favors the higher-$H_0$ region, but does not remove the difficulty in fitting the combined observations.

A stronger preference for $\Lambda$CDM is obtained from PP+DESI and CMB-SPA+PP+DESI, which exclude the SH0ES calibration. In these cases, both the Bayesian evidence and the AIC favor $\Lambda$CDM. The additional parameter in EExpIR neither improves the fit nor supplies the flexibility required by the low-redshift data. Thus, the higher $H_0$ permitted by the CMB-only degeneracies does not translate into a successful joint description of the cosmological observations.

\subsection{Comparison with ExpIR \texorpdfstring{$f(T)$}{} model}

The principal-branch ExpIR $f(T)$ model~\cite{Hashim:2026yoy} has been proposed as a modified-gravity alternative to dark energy, in which the torsion sector effectively behaves as a dynamical dark-energy component evolving entirely within the phantom regime at late times. The EExpIR model considered in the present work extends this framework through the introduction of an additional parameter, $b$. The ExpIR model is recovered as the special case $b=1$, reducing the cosmological parameter space to the same dimensionality as that of the $\Lambda$CDM model. Using the full cosmological dataset, we obtain
\[
b=0.573^{+0.016}_{-0.060},
\]
corresponding to a qualitatively different late-time evolution in which the effective torsional dark energy undergoes a smooth transition from a quintessence-like phase to a phantom regime. In the following, we compare the observational performance of these two teleparallel gravity scenarios.

The principal-branch ExpIR model~\cite{Hashim:2026yoy}, constrained using the CMB-SPA dataset, yields an $H_0$ tension of $0.94\sigma$ relative to SH0ES and $1.65\sigma$ relative to H0DN. Despite this reduction of the tension, the corresponding Bayes factor,
\[
\ln\mathcal{B}_{ij}=-1.19,
\]
indicates a weak preference for $\Lambda$CDM over the ExpIR model. By contrast, the EExpIR model analyzed here leads to larger residual tensions of $2.34\sigma$ and $3.30\sigma$ with SH0ES and H0DN, respectively. Nevertheless, the Bayesian evidence,
\[
\ln\mathcal{B}_{ij}=+0.45,
\]
provides no significant preference between EExpIR and $\Lambda$CDM. These CMB-only results must be interpreted in the context of the parameter correlations and degeneracies that allow different late-time expansion histories to reproduce similar CMB spectra. Low-redshift observations are therefore essential to break these degeneracies and test the expansion histories associated with the higher inferred values of $H_0$.

When the full dataset CMB-SPA+PPS+DESI is considered, the Gaussian estimator gives residual offsets of $1.08$ and $1.90$ from SH0ES and H0DN, respectively, for ExpIR~\cite{Hashim:2026yoy}. These are not independent tension significances, since PPS includes the SH0ES calibration, which also contributes to H0DN. The Bayesian evidence becomes strongly unfavorable,
\[
\ln\mathcal{B}_{ij}=-14.34,
\]
indicating a strong preference for $\Lambda$CDM. For EExpIR, the corresponding residual offsets are $2.18$ and $3.26$, while the Bayes factor,
\[
\ln\mathcal{B}_{ij}=-1.61,
\]
indicates only a weak preference for $\Lambda$CDM for this dataset combination.

These results illustrate the rigidity of the two late-time modified-gravity models. The ExpIR model raises the CMB-inferred $H_0$ substantially, but lacks the freedom to reproduce the low-redshift distance measurements while retaining that higher value. Allowing $b$ to vary in EExpIR provides additional flexibility and reduces the statistical preference for $\Lambda$CDM in the full dataset combination. However, it does not remove the difficulty in fitting the combined observations. The smaller parameter shifts between CMB-only and combined constraints therefore reflect the restricted expansion histories available to these models, rather than improved consistency among the probes. Neither the higher inferred $H_0$ nor the parameter stability alone establishes a successful resolution of the Hubble tension.

\section{Summary and Conclusions}\label{Sec:summary}

In this work we investigated the cosmological viability of the Extended Exponential Infrared (EExpIR) model within the framework of $f(T)$ teleparallel gravity. The model represents a theoretically motivated infrared modification of gravity in which the late-time acceleration of the Universe emerges from torsional geometric effects rather than from an explicit cosmological constant. Owing to the additional parameter governing the infrared deformation, the model is capable of generating a rich dark-energy phenomenology while remaining close to GR at high redshift.

As a first step, we performed a detailed phase-space analysis of the cosmological dynamics. The existence of multiple branches associated with the Lambert-$W$ function leads to qualitatively different evolutionary histories. Among the cases examined, we selected the principal branch with $0.5<b<1$, which approximately reproduces the standard matter-dominated evolution before approaching a stable accelerated attractor. This analysis motivates the branch selection and priors used in the subsequent observational analysis.

We then confronted the model with current cosmological observations, including the CMB-SPA dataset, DESI DR2 BAO measurements, and the PantheonPlus supernova sample. This extends previous background and structure-growth analyses by incorporating the full CMB angular power spectra. While the model provides a good description of the CMB observations, the addition of low-redshift data breaks the CMB parameter degeneracies and exposes the rigidity of its expansion history. The model therefore yields a higher $H_0$ but fits the combined observations less well than $\Lambda$CDM.

Within the adopted prior $0.5<b<1$, the CMB-SPA+PPS+DESI dataset constrains the parameter $b$ to
\[
b = 0.573^{+0.016}_{-0.060},
\]
which corresponds to a quintessence-to-phantom transition in the effective dark-energy equation of state. The corresponding cosmological evolution remains close to $\Lambda$CDM while allowing a nontrivial dynamical dark-energy sector of geometrical origin. Furthermore, the inferred Hubble constant is
\[
H_0 = 70.67^{+0.31}_{-0.35}\ {\rm km\,s^{-1}\,Mpc^{-1}},
\]
which is higher than the corresponding $\Lambda$CDM value obtained from the same dataset. However, the rigidity of the model's low-redshift expansion history prevents this higher $H_0$ from being accompanied by a better description of the combined observations.

At the perturbation level, the model predicts modifications to the effective gravitational coupling and generates a nonvanishing gravitational slip. The torsional contribution to the gravitational slip is most significant on the largest cosmological scales but remains sufficiently suppressed to preserve the good agreement with current CMB observations. Consequently, the temperature and polarization anisotropy spectra are nearly indistinguishable from those of the standard cosmological model over most multipoles, with a mild enhancement of the TT spectrum at $\ell\lesssim5$.

An interesting result of our analysis is that the EExpIR model permits a higher CMB-inferred $H_0$ while preserving a good description of the CMB temperature and polarization spectra. Low-redshift observations provide an essential complementary test by breaking the CMB parameter degeneracies and constraining the late-time expansion history. These observations expose the remaining rigidity of the model: although the additional parameter allows a richer dark-energy evolution, it does not provide sufficient flexibility to fit the combined datasets as well as $\Lambda$CDM.

To quantify the statistical performance of the model, we carried out a Bayesian model-comparison analysis against $\Lambda$CDM. We found no significant preference between the two models when only CMB observations are considered. For the full combination CMB-SPA+PPS+DESI, the Bayesian evidence shows only a weak preference for $\Lambda$CDM. A strong preference for $\Lambda$CDM is instead obtained from PP+DESI and CMB-SPA+PP+DESI. The assessment therefore depends on the dataset combination, with the inclusion of the SH0ES calibration reducing the preference against EExpIR.

The Extended Exponential Infrared $f(T)$ model provides a framework for realizing dynamical dark energy through spacetime torsion, including a smooth quintessence-to-phantom transition. Its ability to preserve the CMB spectra while yielding a higher $H_0$ illustrates the potential of late-time gravitational modifications, while its poorer combined fit identifies the limitations of the particular expansion history considered here. These findings emphasize the importance of testing modified gravity with both CMB and low-redshift observations. Future high-precision measurements of large-scale structure, weak gravitational lensing, redshift-space distortions, and CMB anisotropies will further test the torsional sector and help identify which features of these models can contribute to a successful description of cosmic acceleration.
\begin{acknowledgments}
W.E. and M.H. would like to acknowledge that this paper is based upon work supported by Science, Technology \& Innovation Funding Authority (STDF), Egypt, under Basic Sciences Research Grant (Project ID 53090). J.L.S. would also like to acknowledge funding from ``Xjenza Malta'' as part of the ``Technology Development Programme'' DTP-2024-014 (CosmicLearning) Project. E.D.V. is supported by a Royal Society Dorothy Hodgkin Research Fellowship. This article is based upon work from COST Action CA21136 \emph{Addressing observational tensions in cosmology with systematics and fundamental physics} (CosmoVerse), supported by COST (European Cooperation in Science and Technology).
\end{acknowledgments}



\bibliographystyle{apsrev4-1}
\bibliography{Ref}

\end{document}

%% file: summarytable_LCDM.txt
$\Lambda$CDM  & PP+DESI & CMB-SPA & CMB-SPA+PP+DESI & PPS+DESI & CMB-SPA+PPS+DESI \\
\hline
$\Omega_\text{b} h^2 $ & $ 0.0265^{+0.013}_{-0.0051}$ & $ 0.022399\pm 0.000094$ & $ 0.022466\pm 0.000093$ & $ 0.0286\pm 0.0014$ & $ 0.022516\pm 0.000090$ \\
$\Omega_\text{c} h^2 $ & $ 0.133^{+0.032}_{-0.020}$ & $ 0.12035\pm 0.00095$ & $ 0.11832\pm 0.00061$ & $ 0.1368^{+0.0051}_{-0.0057}$ & $ 0.11770\pm 0.00057$ \\
$H_0 $ & $ 72^{+10}_{-4}$ & $ 67.34\pm 0.39$ & $ 68.17\pm 0.25$ & $ 73.83\pm 0.99$ & $ 68.47\pm 0.23$ \\
$\tau_\text{reio} $ & $--$ & $ 0.0549\pm 0.0054$ & $ 0.0591\pm 0.0054$ & $--$ & $ 0.0611\pm 0.0052$ \\
$\log(10^{10} A_\text{s}) $ & $ ---$ & $ 3.050\pm 0.010$ & $ 3.0596\pm 0.0095$ & $ ---$ & $ 3.0637\pm 0.0095$ \\
$n_\text{s} $ & $ ---$ & $ 0.9693\pm 0.0032$ & $ 0.9734\pm 0.0030$ & $ ---$ & $ 0.9749\pm 0.0028$ \\
\hline
$\Omega_\text{m} $ & $ 0.3046\pm 0.0083$ & $ 0.3162\pm 0.0056$ & $ 0.3043\pm 0.0034$ & $ 0.3045\pm 0.0080$ & $ 0.3005\pm 0.0031$ \\
$\sigma_8 $ & $ 0.78^{+0.16}_{-0.36}$ & $ 0.8150\pm 0.0037$ & $ 0.8133\pm 0.0038$ & $ 0.79^{+0.21}_{-0.36}$ & $ 0.8132\pm 0.0038$ \\
$z_\star $ & $ 1085.371^{+0.044}_{-12}$ & $ 1085.10\pm 0.56$ & $ 1084.45\pm 0.53$ & $ 1079.5^{+1.0}_{-1.2}$ & $ 1084.15\pm 0.51$ \\
$r_\star $ (Mpc) & $ 139.8^{+6.5}_{-17}$ & $ 144.68\pm 0.25$ & $ 145.18\pm 0.18$ & $ 136.3\pm 1.8$ & $ 145.32\pm 0.17$ \\
$100\theta_\star $ & $ 1.052^{+0.029}_{-0.0069}$ & $ 1.04400\pm 0.00041$ & $ 1.04443\pm 0.00039$ & $ 1.0615\pm 0.0052$ & $ 1.04463\pm 0.00038$ \\
$z_\text{re} $ & $ 8.32^{+0.11}_{-2.1}$ & $ 7.73\pm 0.54$ & $ 8.11\pm 0.53$ & $ 7.32\pm 0.22$ & $ 8.28\pm 0.50$ \\
$\sigma_8 (\Omega_\text{m}/0.3)^{0.5} $ & $ 0.78^{+0.16}_{-0.36}$ & $ 0.8367\pm 0.0089$ & $ 0.8192\pm 0.0064$ & $ 0.80^{+0.21}_{-0.37}$ & $ 0.8138\pm 0.0060$ \\
\hline
$\chi^2$ & 1416.17 & 604.17 & 2024.51 & 1465.35 & 2103.85

%% file: summarytable_Model-II.txt
EExpIR  & PP+DESI & CMB-SPA & CMB-SPA+PP+DESI & PPS+DESI & CMB-SPA+PPS+DESI \\
\hline
$\Omega_\text{b} h^2 $ & $ 0.0263^{+0.013}_{-0.0058}$ & $ 0.02243\pm 0.00010$ & $ 0.022444\pm 0.000091$ & $ 0.0256\pm 0.0013$ & $ 0.022482\pm 0.000092$ \\
$\Omega_\text{c} h^2 $ & $ 0.147^{+0.037}_{-0.024}$ & $ 0.1195\pm 0.0011$ & $ 0.11891\pm 0.00062$ & $ 0.1431\pm 0.0056$ & $ 0.11854\pm 0.00064$ \\
$H_0 $ & $ 75^{+10}_{-5}$ & $ 70.32^{+0.52}_{-0.82}$ & $ 70.31^{+0.29}_{-0.33}$ & $ 74.1\pm 1.0$ & $ 70.67^{+0.31}_{-0.35}$ \\
$\tau_\text{reio} $ & $--$ & $ 0.0525\pm 0.0058$ & $ 0.0525\pm 0.0056$ & $--$ & $ 0.0528\pm 0.0057$ \\
$\log(10^{10} A_\text{s}) $ & $ ---$ & $ 3.039\pm 0.012$ & $ 3.038\pm 0.011$ & $ ---$ & $ 3.038\pm 0.012$ \\
$n_\text{s} $ & $ ---$ & $ 0.9713\pm 0.0035$ & $ 0.9727\pm 0.0028$ & $ ---$ & $ 0.9734\pm 0.0029$ \\
$b $ & $ <0.623$ & $ <0.714$ & $ 0.592^{+0.031}_{-0.059}$ & $ <0.620$ & $ 0.573^{+0.016}_{-0.060}$ \\
\hline
$\Omega_\text{m} $ & $ 0.3085\pm 0.0079$ & $ 0.2885^{+0.0077}_{-0.0064}$ & $ 0.2873\pm 0.0034$ & $ 0.3082\pm 0.0082$ & $ 0.2837\pm 0.0033$ \\
$\sigma_8 $ & $ 0.88^{+0.19}_{-0.40}$ & $ 0.8393^{+0.0049}_{-0.0071}$ & $ 0.8378^{+0.0042}_{-0.0054}$ & $ 0.87^{+0.18}_{-0.40}$ & $ 0.8397^{+0.0052}_{-0.0060}$ \\
$z_\star $ & $ 1086.80^{+0.27}_{-12}$ & $ 1085.19\pm 0.58$ & $ 1085.13\pm 0.54$ & $ 1082.8^{+1.2}_{-1.5}$ & $ 1085.02\pm 0.55$ \\
$r_\star $ (Mpc)& $ 136.9^{+6.9}_{-17}$ & $ 144.84\pm 0.27$ & $ 145.00\pm 0.18$ & $ 137.0\pm 1.8$ & $ 145.06\pm 0.19$ \\
$100\theta_\star $ & $ 1.064^{+0.029}_{-0.0067}$ & $ 1.04391\pm 0.00042$ & $ 1.04400\pm 0.00040$ & $ 1.0690\pm 0.0051$ & $ 1.04405\pm 0.00041$ \\
$z_\text{re} $ & $ 8.61^{+0.14}_{-2.2}$ & $ 7.46\pm 0.59$ & $ 7.45^{+0.60}_{-0.54}$ & $ 7.96^{+0.24}_{-0.27}$ & $ 7.46\pm 0.58$ \\
$\sigma_8 (\Omega_\text{m}/0.3)^{0.5} $ & $ 0.90^{+0.20}_{-0.41}$ & $ 0.8229\pm 0.0094$ & $ 0.8198\pm 0.0063$ & $ 0.88^{+0.18}_{-0.40}$ & $ 0.8166\pm 0.0066$ \\
\hline
$\chi^2$ & 1433.44 & 601.92 & 2047.58 & 1482.7 & 2111.12 \\
$\Delta\chi^2$ & 17.27 & $-$2.24 & 23.07 & 17.35 & 7.27 

%% file: Ref.bib
@article{Riess:1998cb,
    author = "Riess, Adam G. and others",
    collaboration = "Supernova Search Team",
    title = "{Observational evidence from supernovae for an accelerating universe and a cosmological constant}",
    eprint = "astro-ph/9805201",
    archivePrefix = "arXiv",
    doi = "10.1086/300499",
    journal = "Astron. J.",
    volume = "116",
    pages = "1009--1038",
    year = "1998"
}

@article{Perlmutter:1998np,
    author = "Perlmutter, S. and others",
    collaboration = "Supernova Cosmology Project",
    title = "{Measurements of $\Omega$ and $\Lambda$ from 42 High Redshift Supernovae}",
    eprint = "astro-ph/9812133",
    archivePrefix = "arXiv",
    reportNumber = "LBNL-41801, LBL-41801",
    doi = "10.1086/307221",
    journal = "Astrophys. J.",
    volume = "517",
    pages = "565--586",
    year = "1999"
}

@article{Planck:2018vyg,
    author = "Aghanim, N. and others",
    collaboration = "Planck",
    title = "{Planck 2018 results. VI. Cosmological parameters}",
    eprint = "1807.06209",
    archivePrefix = "arXiv",
    primaryClass = "astro-ph.CO",
    doi = "10.1051/0004-6361/201833910",
    journal = "Astron. Astrophys.",
    volume = "641",
    pages = "A6",
    year = "2020",
    note = "[Erratum: Astron.Astrophys. 652, C4 (2021)]"
}

@article{Weinberg:1988cp,
    author = "Weinberg, Steven",
    editor = "Hsu, Jong-Ping and Fine, D.",
    title = "{The Cosmological Constant Problem}",
    reportNumber = "UTTG-12-88",
    doi = "10.1103/RevModPhys.61.1",
    journal = "Rev. Mod. Phys.",
    volume = "61",
    pages = "1--23",
    year = "1989"
}

@article{Zlatev:1998tr,
    author = "Zlatev, Ivaylo and Wang, Li-Min and Steinhardt, Paul J.",
    title = "{Quintessence, cosmic coincidence, and the cosmological constant}",
    eprint = "astro-ph/9807002",
    archivePrefix = "arXiv",
    doi = "10.1103/PhysRevLett.82.896",
    journal = "Phys. Rev. Lett.",
    volume = "82",
    pages = "896--899",
    year = "1999"
}

@article{Copeland:2006wr,
    author = "Copeland, Edmund J. and Sami, M. and Tsujikawa, Shinji",
    title = "{Dynamics of dark energy}",
    eprint = "hep-th/0603057",
    archivePrefix = "arXiv",
    doi = "10.1142/S021827180600942X",
    journal = "Int. J. Mod. Phys. D",
    volume = "15",
    pages = "1753--1936",
    year = "2006"
}

@article{Clifton:2011jh,
    author = "Clifton, Timothy and Ferreira, Pedro G. and Padilla, Antonio and Skordis, Constantinos",
    title = "{Modified Gravity and Cosmology}",
    eprint = "1106.2476",
    archivePrefix = "arXiv",
    primaryClass = "astro-ph.CO",
    doi = "10.1016/j.physrep.2012.01.001",
    journal = "Phys. Rept.",
    volume = "513",
    pages = "1--189",
    year = "2012"
}

@article{Bahamonde:2021gfp,
    author = "Bahamonde, Sebastian and Dialektopoulos, Konstantinos F. and Escamilla-Rivera, Celia and Farrugia, Gabriel and Gakis, Viktor and Hendry, Martin and Hohmann, Manuel and Levi Said, Jackson and Mifsud, Jurgen and Di Valentino, Eleonora",
    title = "{Teleparallel gravity: from theory to cosmology}",
    eprint = "2106.13793",
    archivePrefix = "arXiv",
    primaryClass = "gr-qc",
    doi = "10.1088/1361-6633/ac9cef",
    journal = "Rept. Prog. Phys.",
    volume = "86",
    number = "2",
    pages = "026901",
    year = "2023"
}

@article{Bengochea:2008gz,
    author = "Bengochea, Gabriel R. and Ferraro, Rafael",
    title = "{Dark torsion as the cosmic speed-up}",
    eprint = "0812.1205",
    archivePrefix = "arXiv",
    primaryClass = "astro-ph",
    doi = "10.1103/PhysRevD.79.124019",
    journal = "Phys. Rev. D",
    volume = "79",
    pages = "124019",
    year = "2009"
}

@article{Linder:2010py,
    author = "Linder, Eric V.",
    title = "{Einstein's Other Gravity and the Acceleration of the Universe}",
    eprint = "1005.3039",
    archivePrefix = "arXiv",
    primaryClass = "astro-ph.CO",
    doi = "10.1103/PhysRevD.81.127301",
    journal = "Phys. Rev. D",
    volume = "81",
    pages = "127301",
    year = "2010",
    note = "[Erratum: Phys.Rev.D 82, 109902 (2010)]"
}

@article{Santos:2022atq,
    author = "Santos, F. B. M. dos",
    title = "{Updating constraints on phantom crossing f(T) gravity}",
    eprint = "2211.16370",
    archivePrefix = "arXiv",
    primaryClass = "astro-ph.CO",
    doi = "10.1088/1475-7516/2023/06/039",
    journal = "JCAP",
    volume = "06",
    pages = "039",
    year = "2023"
}

@article{Scolnic:2021amr,
    author = "Scolnic, Dan and others",
    title = "{The Pantheon+ Analysis: The Full Data Set and Light-curve Release}",
    eprint = "2112.03863",
    archivePrefix = "arXiv",
    primaryClass = "astro-ph.CO",
    doi = "10.3847/1538-4357/ac8b7a",
    journal = "Astrophys. J.",
    volume = "938",
    number = "2",
    pages = "113",
    year = "2022"
}

@article{CosmoVerseNetwork:2025alb,
    author = "Di Valentino, Eleonora and others",
    collaboration = "CosmoVerse Network",
    title = "{The CosmoVerse White Paper: Addressing observational tensions in cosmology with systematics and fundamental physics}",
    eprint = "2504.01669",
    archivePrefix = "arXiv",
    primaryClass = "astro-ph.CO",
    doi = "10.1016/j.dark.2025.101965",
    journal = "Phys. Dark Univ.",
    volume = "49",
    pages = "101965",
    year = "2025"
}

@article{Krssak:2015oua,
    author = "Kr{\v{s}}{\v{s}}{\'a}k, Martin and Saridakis, Emmanuel N.",
    title = "{The covariant formulation of f(T) gravity}",
    eprint = "1510.08432",
    archivePrefix = "arXiv",
    primaryClass = "gr-qc",
    doi = "10.1088/0264-9381/33/11/115009",
    journal = "Class. Quant. Grav.",
    volume = "33",
    number = "11",
    pages = "115009",
    year = "2016"
}

@book{Weitzenbock:1923efa,
    author = {Weitzenb{\"o}ck, Roland},
    title = "{Invariantentheorie}",
    publisher = "P. Noordhoff",
    address = "Groningen",
    year = "1923"
}

@article{Maluf:2013gaa,
    author = "Maluf, J. W.",
    title = "{The teleparallel equivalent of general relativity}",
    eprint = "1303.3897",
    archivePrefix = "arXiv",
    primaryClass = "gr-qc",
    doi = "10.1002/andp.201200272",
    journal = "Annalen Phys.",
    volume = "525",
    pages = "339--357",
    year = "2013"
}

@article{Li:2010cg,
    author = "Li, Baojiu and Sotiriou, Thomas P. and Barrow, John D.",
    title = "{$f(T)$ gravity and local Lorentz invariance}",
    eprint = "1010.1041",
    archivePrefix = "arXiv",
    primaryClass = "gr-qc",
    doi = "10.1103/PhysRevD.83.064035",
    journal = "Phys. Rev. D",
    volume = "83",
    pages = "064035",
    year = "2011"
}

@article{Awad:2017yod,
    author = "Awad, A. and El Hanafy, W. and Nashed, G. G. L. and Saridakis, Emmanuel N.",
    title = "{Phase Portraits of general f(T) Cosmology}",
    eprint = "1710.10194",
    archivePrefix = "arXiv",
    primaryClass = "gr-qc",
    doi = "10.1088/1475-7516/2018/02/052",
    journal = "JCAP",
    volume = "02",
    pages = "052",
    year = "2018"
}

@article{Golovnev:2018wbh,
    author = "Golovnev, Alexey and Koivisto, Tomi",
    title = "{Cosmological perturbations in modified teleparallel gravity models}",
    eprint = "1808.05565",
    archivePrefix = "arXiv",
    primaryClass = "gr-qc",
    reportNumber = "NORDITA 2018-073",
    doi = "10.1088/1475-7516/2018/11/012",
    journal = "JCAP",
    volume = "11",
    pages = "012",
    year = "2018"
}

@article{Bertschinger:2008zb,
    author = "Bertschinger, Edmund and Zukin, Phillip",
    title = "{Distinguishing Modified Gravity from Dark Energy}",
    eprint = "0801.2431",
    archivePrefix = "arXiv",
    primaryClass = "astro-ph",
    doi = "10.1103/PhysRevD.78.024015",
    journal = "Phys. Rev. D",
    volume = "78",
    pages = "024015",
    year = "2008"
}

@article{Bean:2010zq,
    author = "Bean, Rachel and Tangmatitham, Matipon",
    title = "{Current constraints on the cosmic growth history}",
    eprint = "1002.4197",
    archivePrefix = "arXiv",
    primaryClass = "astro-ph.CO",
    doi = "10.1103/PhysRevD.81.083534",
    journal = "Phys. Rev. D",
    volume = "81",
    pages = "083534",
    year = "2010"
}

@article{Ma:1995ey,
    author = "Ma, Chung-Pei and Bertschinger, Edmund",
    title = "{Cosmological perturbation theory in the synchronous and conformal Newtonian gauges}",
    eprint = "astro-ph/9506072",
    archivePrefix = "arXiv",
    doi = "10.1086/176550",
    journal = "Astrophys. J.",
    volume = "455",
    pages = "7--25",
    year = "1995"
}

@article{DESI:2025zgx,
    author = "Abdul Karim, M. and others",
    collaboration = "DESI",
    title = "{DESI DR2 results. II. Measurements of baryon acoustic oscillations and cosmological constraints}",
    eprint = "2503.14738",
    archivePrefix = "arXiv",
    primaryClass = "astro-ph.CO",
    reportNumber = "FERMILAB-PUB-25-0169-PPD",
    doi = "10.1103/tr6y-kpc6",
    journal = "Phys. Rev. D",
    volume = "112",
    number = "8",
    pages = "083515",
    year = "2025"
}

@article{Hashim:2020sez,
    author = "Hashim, Mahmoud and El Hanafy, Waleed and Golovnev, Alexey and El-Zant, Amr A.",
    title = "{Toward a concordance teleparallel cosmology. Part I. Background dynamics}",
    eprint = "2010.14964",
    archivePrefix = "arXiv",
    primaryClass = "astro-ph.CO",
    doi = "10.1088/1475-7516/2021/07/052",
    journal = "JCAP",
    volume = "07",
    pages = "052",
    year = "2021"
}

@article{Hashim:2021pkq,
    author = "Hashim, Mahmoud and El-Zant, Amr A. and El Hanafy, Waleed and Golovnev, Alexey",
    title = "{Toward a concordance teleparallel cosmology. Part~II. Linear perturbation}",
    eprint = "2104.08311",
    archivePrefix = "arXiv",
    primaryClass = "astro-ph.CO",
    doi = "10.1088/1475-7516/2021/07/053",
    journal = "JCAP",
    volume = "07",
    pages = "053",
    year = "2021"
}

@article{Akarsu:2024nas,
    author = "Akarsu, Ozgur and Bulduk, Bilal and De Felice, Antonio and Kat{\i}rc{\i}, Nihan and Uzun, N. Merve",
    title = "{Unexplored regions in teleparallel f(T) gravity: Sign-changing dark energy density}",
    eprint = "2410.23068",
    archivePrefix = "arXiv",
    primaryClass = "gr-qc",
    reportNumber = "YITP-24-119",
    doi = "10.1103/1xd4-k91h",
    journal = "Phys. Rev. D",
    volume = "112",
    number = "8",
    pages = "083532",
    year = "2025"
}

@article{Hashim:2026yoy,
    author = "Hashim, Mahmoud and Di Valentino, Eleonora and Levi Said, Jackson and El Hanafy, Waleed",
    title = "{Cosmological Viability of Exponential Infrared $f(T)$ Gravity}",
    eprint = "2606.31324",
    archivePrefix = "arXiv",
    primaryClass = "astro-ph.CO",
    month = "6",
    year = "2026"
}

@article{Bedair:2026zvn,
    author = "Bedair, Mohamed and Hashim, Mahmoud and Levi Said, Jackson and El Hanafy, Waleed",
    title = "{A Six-Parameter Teleparallel Cosmology Beyond $Λ$CDM: Sign-Changing Torsional Dark Energy and Implications for $H_0$}",
    eprint = "2609.10560",
    archivePrefix = "arXiv",
    primaryClass = "physics.gen-ph",
    month = "8",
    year = "2026"
}

@article{Awad:2013tha,
    author = "Awad, Adel",
    title = "{Fixed points and FLRW cosmologies: Flat case}",
    eprint = "1303.2014",
    archivePrefix = "arXiv",
    primaryClass = "gr-qc",
    doi = "10.1103/PhysRevD.87.103001",
    journal = "Phys. Rev. D",
    volume = "87",
    number = "10",
    pages = "103001",
    year = "2013",
    note = "[Erratum: Phys.Rev.D 87, 109902 (2013)]"
}

@article{El-Zant:2018bsc,
    author = "El-Zant, Amr and El Hanafy, Waleed and Elgammal, Sherif",
    title = "{$H_0$ Tension and the Phantom Regime: A Case Study in Terms of an Infrared $f(T)$ Gravity}",
    eprint = "1809.09390",
    archivePrefix = "arXiv",
    primaryClass = "gr-qc",
    doi = "10.3847/1538-4357/aafa12",
    journal = "Astrophys. J.",
    volume = "871",
    number = "2",
    pages = "210",
    year = "2019"
}

@article{Torrado:2020dgo,
    author = "Torrado, Jesus and Lewis, Antony",
    title = "{Cobaya: Code for Bayesian Analysis of hierarchical physical models}",
    eprint = "2005.05290",
    archivePrefix = "arXiv",
    primaryClass = "astro-ph.IM",
    reportNumber = "TTK-20-15",
    doi = "10.1088/1475-7516/2021/05/057",
    journal = "JCAP",
    volume = "05",
    pages = "057",
    year = "2021"
}

@article{Blas:2011rf,
    author = "Blas, Diego and Lesgourgues, Julien and Tram, Thomas",
    title = "{The Cosmic Linear Anisotropy Solving System (CLASS) II: Approximation schemes}",
    eprint = "1104.2933",
    archivePrefix = "arXiv",
    primaryClass = "astro-ph.CO",
    reportNumber = "CERN-PH-TH-2011-082, LAPTH-010-11",
    doi = "10.1088/1475-7516/2011/07/034",
    journal = "JCAP",
    volume = "07",
    pages = "034",
    year = "2011"
}

@article{Gelman:1992zz,
    author = "Gelman, Andrew and Rubin, Donald B.",
    title = "{Inference from Iterative Simulation Using Multiple Sequences}",
    doi = "10.1214/ss/1177011136",
    journal = "Statist. Sci.",
    volume = "7",
    pages = "457--472",
    year = "1992"
}

@article{Planck:2019nip,
    author = "Aghanim, N. and others",
    collaboration = "Planck",
    title = "{Planck 2018 results. V. CMB power spectra and likelihoods}",
    eprint = "1907.12875",
    archivePrefix = "arXiv",
    primaryClass = "astro-ph.CO",
    doi = "10.1051/0004-6361/201936386",
    journal = "Astron. Astrophys.",
    volume = "641",
    pages = "A5",
    year = "2020"
}

@article{AtacamaCosmologyTelescope:2025blo,
    author = "Louis, Thibaut and others",
    collaboration = "Atacama Cosmology Telescope",
    title = "{The Atacama Cosmology Telescope: DR6 power spectra, likelihoods and {\ensuremath{\Lambda}}CDM parameters}",
    eprint = "2503.14452",
    archivePrefix = "arXiv",
    primaryClass = "astro-ph.CO",
    reportNumber = "FERMILAB-PUB-25-0071-PPD",
    doi = "10.1088/1475-7516/2025/11/062",
    journal = "JCAP",
    volume = "11",
    pages = "062",
    year = "2025"
}

@article{Carron:2022eyg,
    author = "Carron, Julien and Mirmelstein, Mark and Lewis, Antony",
    title = "{CMB lensing from Planck PR4~maps}",
    eprint = "2206.07773",
    archivePrefix = "arXiv",
    primaryClass = "astro-ph.CO",
    doi = "10.1088/1475-7516/2022/09/039",
    journal = "JCAP",
    volume = "09",
    pages = "039",
    year = "2022"
}

@article{ACT:2023dou,
    author = "Qu, Frank J. and others",
    collaboration = "ACT",
    title = "{The Atacama Cosmology Telescope: A Measurement of the DR6 CMB Lensing Power Spectrum and Its Implications for Structure Growth}",
    eprint = "2304.05202",
    archivePrefix = "arXiv",
    primaryClass = "astro-ph.CO",
    reportNumber = "FERMILAB-PUB-23-237-PPD, FERMILAB-PUB-23-237-PPD",
    doi = "10.3847/1538-4357/acfe06",
    journal = "Astrophys. J.",
    volume = "962",
    number = "2",
    pages = "112",
    year = "2024"
}

@article{ACT:2023kun,
    author = "Madhavacheril, Mathew S. and others",
    collaboration = "ACT",
    title = "{The Atacama Cosmology Telescope: DR6 Gravitational Lensing Map and Cosmological Parameters}",
    eprint = "2304.05203",
    archivePrefix = "arXiv",
    primaryClass = "astro-ph.CO",
    reportNumber = "FERMILAB-PUB-23-206-PPD",
    doi = "10.3847/1538-4357/acff5f",
    journal = "Astrophys. J.",
    volume = "962",
    number = "2",
    pages = "113",
    year = "2024"
}

@article{SPT-3G:2025bzu,
    author = "Camphuis, E. and others",
    collaboration = "SPT-3G",
    title = "{SPT-3G D1: CMB temperature and polarization power spectra and cosmology from 2019 and 2020 observations of the SPT-3G main field}",
    eprint = "2506.20707",
    archivePrefix = "arXiv",
    primaryClass = "astro-ph.CO",
    reportNumber = "FERMILAB-PUB-25-0144-PPD",
    doi = "10.1103/7wt3-9v2y",
    journal = "Phys. Rev. D",
    volume = "113",
    number = "8",
    pages = "083504",
    year = "2026"
}

@article{SPT-3G:2024atg,
    author = "Ge, F. and others",
    collaboration = "SPT-3G",
    title = "{Cosmology from CMB lensing and delensed EE power spectra using 2019{\textendash}2020 SPT-3G polarization data}",
    eprint = "2411.06000",
    archivePrefix = "arXiv",
    primaryClass = "astro-ph.CO",
    reportNumber = "FERMILAB-PUB-24-0840-PPD",
    doi = "10.1103/PhysRevD.111.083534",
    journal = "Phys. Rev. D",
    volume = "111",
    number = "8",
    pages = "083534",
    year = "2025"
}

@article{Brout:2022vxf,
    author = "Brout, Dillon and others",
    title = "{The Pantheon+ Analysis: Cosmological Constraints}",
    eprint = "2202.04077",
    archivePrefix = "arXiv",
    primaryClass = "astro-ph.CO",
    doi = "10.3847/1538-4357/ac8e04",
    journal = "Astrophys. J.",
    volume = "938",
    number = "2",
    pages = "110",
    year = "2022"
}

@article{Gorbunov:2026sly,
    author = "Gorbunov, Dmitry and Nedelko, Nikita",
    title = "{Changing constraints on {\ensuremath{\sum}}m{\ensuremath{\nu}} from SPT-3G 2018 to D1 in the context of DESI DR2}",
    eprint = "2601.16277",
    archivePrefix = "arXiv",
    primaryClass = "astro-ph.CO",
    doi = "10.1016/j.physletb.2026.140659",
    journal = "Phys. Lett. B",
    volume = "879",
    pages = "140659",
    year = "2026"
}

@article{Lemos:2018smw,
    author = "Lemos, Pablo and Lee, Elizabeth and Efstathiou, George and Gratton, Steven",
    title = "{Model independent $H(z)$ reconstruction using the cosmic inverse distance ladder}",
    eprint = "1806.06781",
    archivePrefix = "arXiv",
    primaryClass = "astro-ph.CO",
    doi = "10.1093/mnras/sty3082",
    journal = "Mon. Not. Roy. Astron. Soc.",
    volume = "483",
    number = "4",
    pages = "4803--4810",
    year = "2019"
}

@article{Riess:2021jrx,
    author = "Riess, Adam G. and others",
    title = "{A Comprehensive Measurement of the Local Value of the Hubble Constant with 1 km s$^{−1}$ Mpc$^{−1}$ Uncertainty from the Hubble Space Telescope and the SH0ES Team}",
    eprint = "2112.04510",
    archivePrefix = "arXiv",
    primaryClass = "astro-ph.CO",
    doi = "10.3847/2041-8213/ac5c5b",
    journal = "Astrophys. J. Lett.",
    volume = "934",
    number = "1",
    pages = "L7",
    year = "2022"
}

@article{H0DN:2025lyy,
    author = "Casertano, Stefano and others",
    collaboration = "H0DN",
    title = "{The Local Distance Network: A community consensus report on the measurement of the Hubble constant at {\ensuremath{\sim}}1{\%} precision}",
    eprint = "2510.23823",
    archivePrefix = "arXiv",
    primaryClass = "astro-ph.CO",
    doi = "10.1051/0004-6361/202557993",
    journal = "Astron. Astrophys.",
    volume = "708",
    pages = "A166",
    year = "2026"
}

@article{Akaike:1974vps,
    author = "Akaike, H.",
    title = "{A new look at the statistical model identification}",
    doi = "10.1109/TAC.1974.1100705",
    journal = "IEEE Trans. Automatic Control",
    volume = "19",
    number = "6",
    pages = "716--723",
    year = "1974"
}

@article{Liddle:2007fy,
    author = "Liddle, Andrew R",
    title = "{Information criteria for astrophysical model selection}",
    eprint = "astro-ph/0701113",
    archivePrefix = "arXiv",
    doi = "10.1111/j.1745-3933.2007.00306.x",
    journal = "Mon. Not. Roy. Astron. Soc.",
    volume = "377",
    pages = "L74--L78",
    year = "2007"
}

@article{Trotta:2008qt,
    author = "Trotta, Roberto",
    title = "{Bayes in the sky: Bayesian inference and model selection in cosmology}",
    eprint = "0803.4089",
    archivePrefix = "arXiv",
    primaryClass = "astro-ph",
    doi = "10.1080/00107510802066753",
    journal = "Contemp. Phys.",
    volume = "49",
    pages = "71--104",
    year = "2008"
}

@article{harmonic,
    author = {McEwen, Jason D. and Wallis, Christopher G. R.
              and Price, Matthew A. and Docherty, Matthew M.},
    title = {Machine learning assisted Bayesian model comparison:
             learnt harmonic mean estimator},
    eprint = {2111.12720},
    archivePrefix = {arXiv},
    primaryClass = {stat.ME},
    year = {2021}
}

@article{Polanska:2024arc,
    author = {Polanska, Alicja and Price, Matthew A. and
              Piras, Davide and Spurio Mancini, Alessio and
              McEwen, Jason D.},
    title = {Learned harmonic mean estimation of the Bayesian
             evidence with normalizing flows},
    journal = {The Open Journal of Astrophysics},
    volume = {8},
    year = {2025},
    doi = {10.33232/001c.146026},
    eprint = {2405.05969},
    archivePrefix = {arXiv},
    primaryClass = {astro-ph.IM}
}

@article{DES:2020hen,
    author = "Lemos, P. and others",
    collaboration = "DES",
    title = "{Assessing tension metrics with dark energy survey and Planck data}",
    eprint = "2012.09554",
    archivePrefix = "arXiv",
    primaryClass = "astro-ph.CO",
    reportNumber = "FERMILAB-PUB-20-662-AE",
    doi = "10.1093/mnras/stab1670",
    journal = "Mon. Not. Roy. Astron. Soc.",
    volume = "505",
    number = "4",
    pages = "6179--6194",
    year = "2021"
}
